\documentclass[12pt,a4paper,dvips]{article}
\usepackage{a4p}
\usepackage{cite,mcite}
\usepackage{graphicx}

\newcommand{\ra}{\mbox{$\rightarrow$}}

\begin{document}
\begin{titlepage}
\def\thefootnote{\fnsymbol{footnote}}       

\begin{center}
\mbox{ } 

\vspace*{-1cm}

{\Large \mbox{\hspace{-1.3cm} EUROPEAN ORGANIZATION FOR NUCLEAR RESEARCH}} 

\end{center}
\vskip 0.5cm
\begin{flushright}
\large
hep-ph/9712283 \\
IEKP-KA/97-14 \\
December 6, 1997
\end{flushright}
\begin{center}
\vskip 0.5cm
{\Huge\bf
Searches for Higgs Bosons at LEP2}
\vskip 1.0cm
{\Large\bf Andr\'e Sopczak}\\
\smallskip
\large University of Karlsruhe

\vskip 2.5cm
\centerline{\large \bf Abstract}
\end{center}

\renewcommand{\baselinestretch} {1.2}

\large
The latest results of Higgs boson searches from the four LEP experiments,
ALEPH, DELPHI, L3 and OPAL, are reviewed using the data taken in 1996 at
center-of-mass energies between 161 and 172~GeV.
No signal was observed.
The 95\% CL combined lower mass limit for the Minimal Standard Model 
(MSM) Higgs boson has increased from 66~GeV at LEP1 to 77.5~GeV 
with the first LEP2 data. In the framework of the
Two Higgs Doublet Model, the charged Higgs boson mass limit 
has increased from 44~GeV to 54.5~GeV, independent
of the decay branching ratio. Large new ($m_h,\tan\beta$) and
($m_h,m_A$) parameter regions are excluded in the framework of the
Minimal Supersymmetric extension of the Standard Model (MSSM). 
Preliminary results from the 1997 data-taking at 183~GeV are presented,
and the prospects for a discovery in the near future are given.
\renewcommand{\baselinestretch} {1.}

\normalsize
\vspace{2cm}
\begin{center}
{\sl
Presented at the 1st Int. Workshop on Non-Accelerator Physics, 
Dubna, July 1997, 
to be published in the proceedings, and at the Aspen Center for Physics, 
August 1997.
}
\end{center}
\vfill
\end{titlepage}

\newpage
\thispagestyle{empty}
\mbox{ }
\newpage
\setcounter{page}{1}
\begin{center}
{\bf \large Searches for Higgs Bosons at LEP2}
\\
{\bf Andr\'e Sopczak}\footnote{\normalsize e-mail: andre.sopczak@cern.ch}
\\
{Karlsruhe University}
\end{center}

\begin{abstract}
The latest results of Higgs boson searches from the four LEP experiments,
ALEPH, DELPHI, L3 and OPAL, are reviewed using the data taken in 1996 at
center-of-mass energies between 161 and 172~GeV.
No signal was observed.
The 95\% CL combined lower mass limit for the Minimal Standard Model (MSM) 
Higgs boson has increased from 66~GeV at LEP1 to 77.5~GeV
with the first LEP2 data. In the framework of the
Two Higgs Doublet Model, the charged Higgs boson mass limit 
has increased from 44~GeV to 54.5~GeV, independent
of the decay branching ratio. Large new ($m_h,\tan\beta$) and
($m_h,m_A$) parameter regions are excluded in the framework of the
Minimal Supersymmetric extension of the Standard Model (MSSM). 
Preliminary results from the 1997 data-taking at 183~GeV are presented,
and the prospects for a discovery in the near future are given.

\end{abstract}

\section{Introduction}

After six years of data-taking at the Z resonance (LEP1), the LEP machine
energy was increased (LEP2), firstly to 130~GeV in fall 1995, and 
successively to 172~GeV in 1996 and then to 183~GeV in 1997.
This review focuses on the results based on the data collected at $\sqrt{s}%
=161 $ to 172~GeV in 1996 with a total luminosity of about 21~pb$^{-1}$ for
each LEP experiment. At 183~GeV, LEP experiments have collected data of about 
55~pb$^{-1}$ each.
For example, first LEP2 results were summarized in~\cite{eilam,bill,patrick},
and final results from LEP1 were reviewed in~\cite{as97}.

The experimental evidence of Higgs bosons is crucial for understanding 
the mechanism of SU(2)$\times$U(1) symmetry breaking and mass generation 
in gauge theories. The Higgs mass is not predicted by the theory.

\begin{figure}[hbtp]
\begin{minipage}{0.48\textwidth}
  \begin{center}
    \includegraphics[width=\textwidth]{mwmt.eps}
  \end{center}
\vspace*{-0.7cm}
  \caption[]{\label{fig:hmass} Measured $t$ and $W$ masses with error
             ellipses. In the MSM (grey region) light Higgs boson masses
             are favored.
             The hatched area shows the prediction in the MSSM.}
\end{minipage}
\hfill
\begin{minipage}{0.48\textwidth}
\vspace*{-0.9cm}
  \begin{center}
    \includegraphics[width=\textwidth]{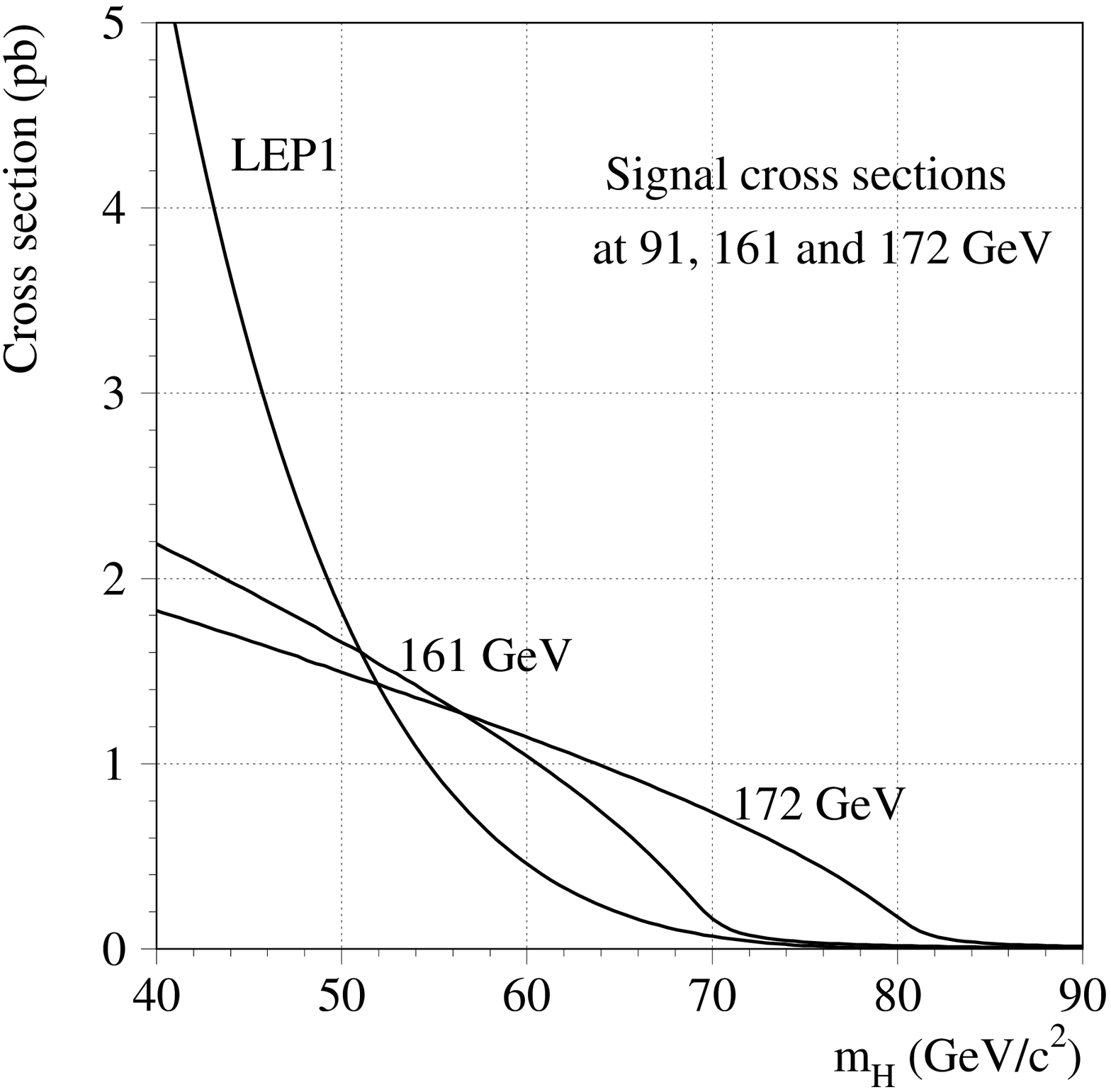}
  \end{center}
\vspace*{-0.6cm}
  \caption[]{\label{fig:xsec}
             Cross section for the MSM Higgs
             production $e^+e^- \rightarrow ZH$
             at different center-of-mass energies.}
\end{minipage}
\end{figure}

The progress made at  Tevatron and LEP with respect to precise measurements of 
electroweak observables, mainly $t$-quark and $W$-boson masses, has led to improved 
indirect constraints on the Higgs boson mass.
Figure~\ref{fig:hmass} (from~\cite{karlsruhe,lepew}) shows a comparison of 
$t$ and $W$ measurements and theoretical predictions for different MSM 
Higgs boson masses. Light Higgs boson masses are favored and an upper 
mass limit of 295~GeV at 95\% CL is derived~\cite{lepew,jens}. In addition, 
a typical mass region in the MSSM is shown, which predicts larger $W$ masses. 
The current precision does not allow a distinction to be made 
between the MSM and the MSSM, leaving the MSSM as an attractive extension of the MSM.

There are important differences between the Higgs boson searches at LEP1 and
LEP2:

\begin{itemize}
\item  The signal-to-background ratio is much better at LEP2. For example,
at LEP1 the ratio of background to the expected number of signal events
for a 60~GeV Higgs boson was about 50,000, while at LEP2 the ratio is 
about 200.
The large background rate at LEP1 required a very detailed simulation of
detector effects and rare background reactions.
Furthermore, the dominant hadronic 
Higgs boson signature ($HZ\rightarrow qq qq$) was useless at LEP1 because of
the overwhelming QCD background.

\item  In addition to the larger numbers of distinguished search signatures
at LEP2, the different center-of-mass energies are treated as different
channels with separate signal and background simulations. The different MSM
Higgs boson searches at 161 and 172~GeV center-of-mass energies amount to 12
search channels. The statistical treatment of the search optimization is
therefore more complex than for LEP1.

\item  While the expected MSM Higgs production at LEP1 involved a real $Z$
decaying into a Higgs boson and a very virtual $Z$, at LEP2 the Higgs boson
could be produced in association with an on-shell $Z$. This additional
information about the final state $Z$ boson mass gives rise to better 
Higgs boson mass reconstruction, and thus greater sensitivity for a Higgs 
boson signal because of better background rejection.

\item  At LEP1 almost no irreducible background was expected, while at LEP2
some processes, which are now kinematically allowed, lead to 
signatures identical to those expected for the signal. 
Most important, when Higgs and $Z$ masses are 
almost degenerate, $ZZ$ background where one $Z$ decays into $bb$, 
is not distinguishable from a Higgs signal.
\end{itemize}

This article is structured as follows.
In Section~\ref{sec:msm} the MSM Higgs results are compared, and 
in Section~\ref{sec:msmcombi} the combination
of these results with significantly greater sensitivity is reviewed. In 
Section~\ref{sec:thdm} the results of searches in the Two Higgs Doublet
model are summarized, in Section~\ref{sec:mssm} interpretations in the MSSM 
are given, and Section~\ref{sec:outlook} contains a brief outlook.

\section{\label{sec:msm}MSM Higgs Boson Results}

For the MSM Higgs boson search, only the LEP2 data taken at the highest
center-of-mass energy are relevant. Figure~\ref{fig:xsec} (from~\cite{aleph}) 
shows the Higgs boson production cross sections for LEP1 
at $\sqrt{s}=91$, and LEP2 at $\sqrt{s}=161$ and 172~GeV. 
For Higgs boson masses above about 70~GeV, the 172~GeV data dominates the 
production cross section. 
The Higgs and $Z$ decay modes determine the event signature in the detector.
The most important characteristic is that Higgs decays predominantly
($\sim 86$\%) into $bb$. Each experiment has performed searches 
in the final states listed in Table~\ref{tab:br}.
\begin{table}[hbt]
\begin{minipage}{0.60\textwidth}
\begin{tabular} {|c|c|}\hline
Final states                 &   Branching fraction (in \%) \\ \hline
$(Z\ra qq)(H \ra qq)$        &   64    \\
$(Z\ra \nu\nu)(H \ra qq)$        &   18    \\
$(Z\ra ee, \mu\mu)(H \ra qq)$&    6.2  \\
$(Z\ra\tau\tau)(H\ra qq)$    &    3.1  \\
$(Z\ra qq)(H\ra\tau\tau)$    &    5.4 \\\hline
\end{tabular}
\end{minipage}
\hfill
\begin{minipage}{0.39\textwidth}
\caption{\label{tab:br} 
         Final-state particles in the analyzed Higgs channels and 
         approximative branching fractions.}
\end{minipage}
\end{table}

\begin{figure}[htbp]
  \begin{center}
    \includegraphics[width=\textwidth]{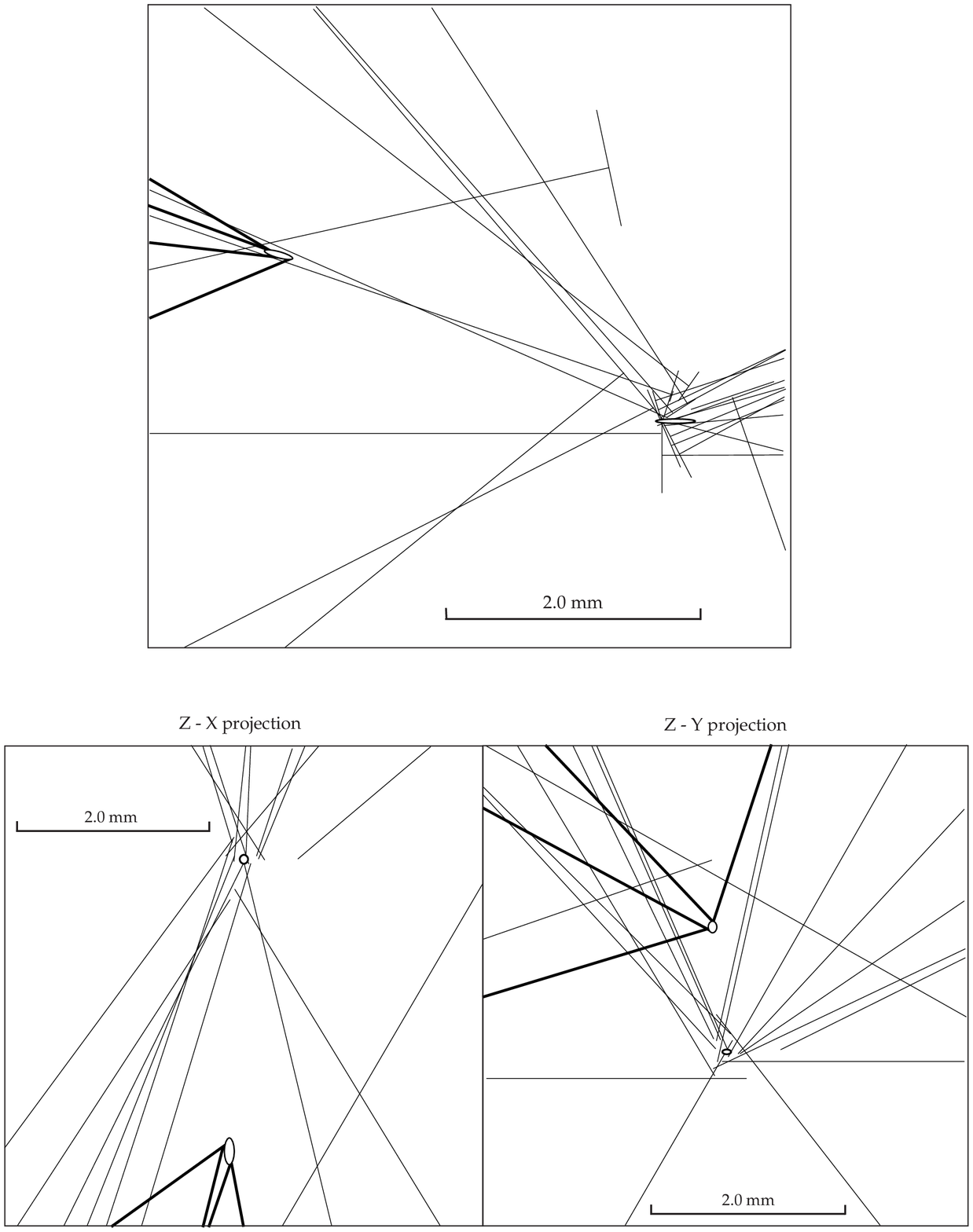}
  \end{center}
  \caption[]{\label{fig:bbtag} DELPHI tagging of $b$-quark jets
             for the example of the $H\nu\nu$ candidate.  
             Primary and secondary vertices are shown in three dimensions.
             The bold-faced tracks define the secondary vertex.}
\end{figure}

The four LEP experiments have chosen different event selection strategies 
with respect to the number of expected background events.
ALEPH uses a very tight event selection, such that less than one background
event is expected. No candidate event is observed in their data. DELPHI
expects about four background events and observes two candidates. 
OPAL expects about four background events and also observes two candidates. 
Very loose cuts are applied by L3, where approximately ten background 
events are expected and six data events pass their selection for any mass
hypothesis between 60 and 70~GeV.
L3 uses different selection cuts for different 
mass regions; thus their total number of candidate events is 33,
consistent with 38 expected background events.
The two most important channels $Hqq$ and $H\nu \nu$, with the largest 
expected event rates, are discussed:
\begin{itemize}
\item
The $Hqq$ channel has a four-jet event topology where one jet pair
originates from hadronic $Z$ decay, and the other from the Higgs decay. 
The Higgs boson decay branching fraction into a $b$-quark pair is about
85\%; therefore, 
the search uses $b$-quark tagging to reduce the background from 
$WW\rightarrow qqqq$, and QCD background where gluon emission leads 
to multiple-jet events. Secondary vertices arise in $b$-quark events 
because of the production of long-lived $B$-mesons. 

\item
In the $H\nu \nu$ channel, events are characterized by two acoplanar jets
carrying $b$-flavor and large missing mass compatible with the $Z$ mass.
Background are $qq$ events where either one jet is mismeasured or an
energetic neutrino is produced in a semileptonic decay. Other
reactions leading to missing energy are $WW\rightarrow l\nu qq$ and $We\nu
\rightarrow qqe\nu $ where the charged lepton escaped detection. 
For an efficient $b$-quark tagging, secondary 
vertices are reconstructed in three dimensions as shown for example
in Fig.~\ref{fig:bbtag} (from~\cite{delphi}).
In general,
events with undetected particles along the beam pipe from two-photon events
and hard initial photon radiation lead to the missing energy signature.
Dedicated cuts for each process reduce such background while maintaining high
selection efficiencies. 
\end{itemize}

\begin{figure}[htbp]
  \begin{center}
    \includegraphics[width=0.49\textwidth]{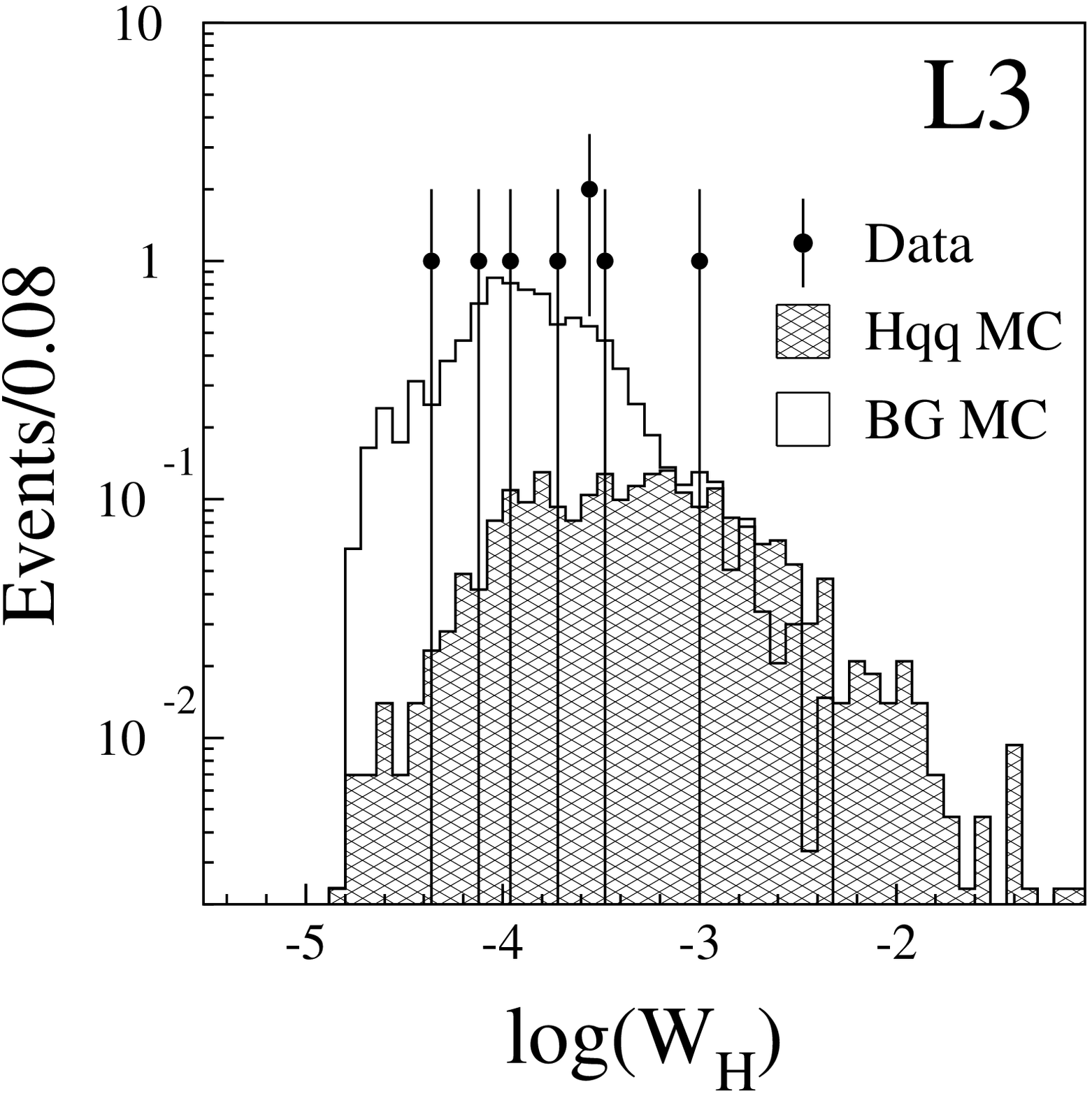}
    \includegraphics[width=0.49\textwidth]{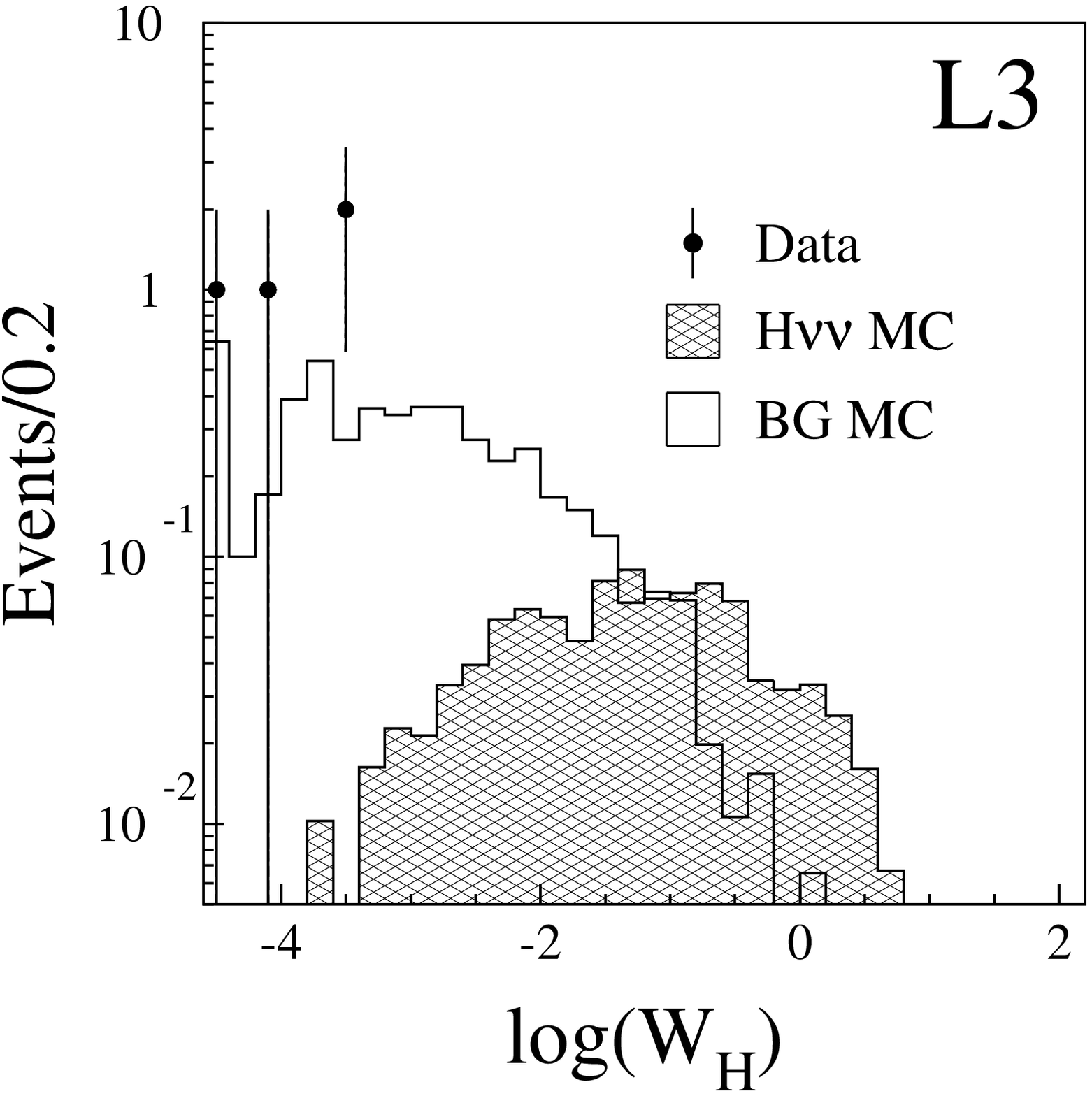}
  \end{center}
\vspace*{-0.5cm}
\caption[]{\label{fig:l3can} 
           L3 candidates in the $Hqq$ channel (left) and the $H\nu\nu$ 
           channel (right). The weights of the candidates 
           are shown compared to background (open histogram) and signal 
           (hatched histogram).}
\end{figure}

Table~\ref{tab:selection}
(from~\cite{lepwg} based on~\cite{aleph,delphi,l3,opal})
gives the expected background, the simulated detection efficiencies 
and the numbers of expected signal events for all channels.
Candidate events are listed for DELPHI and OPAL in Table~\ref{tab:delopal} 
and are given for L3 in Fig.~\ref{fig:l3can}. 
No indication of a signal is found in the reconstructed mass
spectrum. L3 attributes a weight to each candidate and the remaining
candidates are more background- than signal-like.

\begin{figure}[H]
\begin{minipage}{0.48\textwidth}
\vspace*{-1cm}
  \begin{center}
    \includegraphics[width=\textwidth]{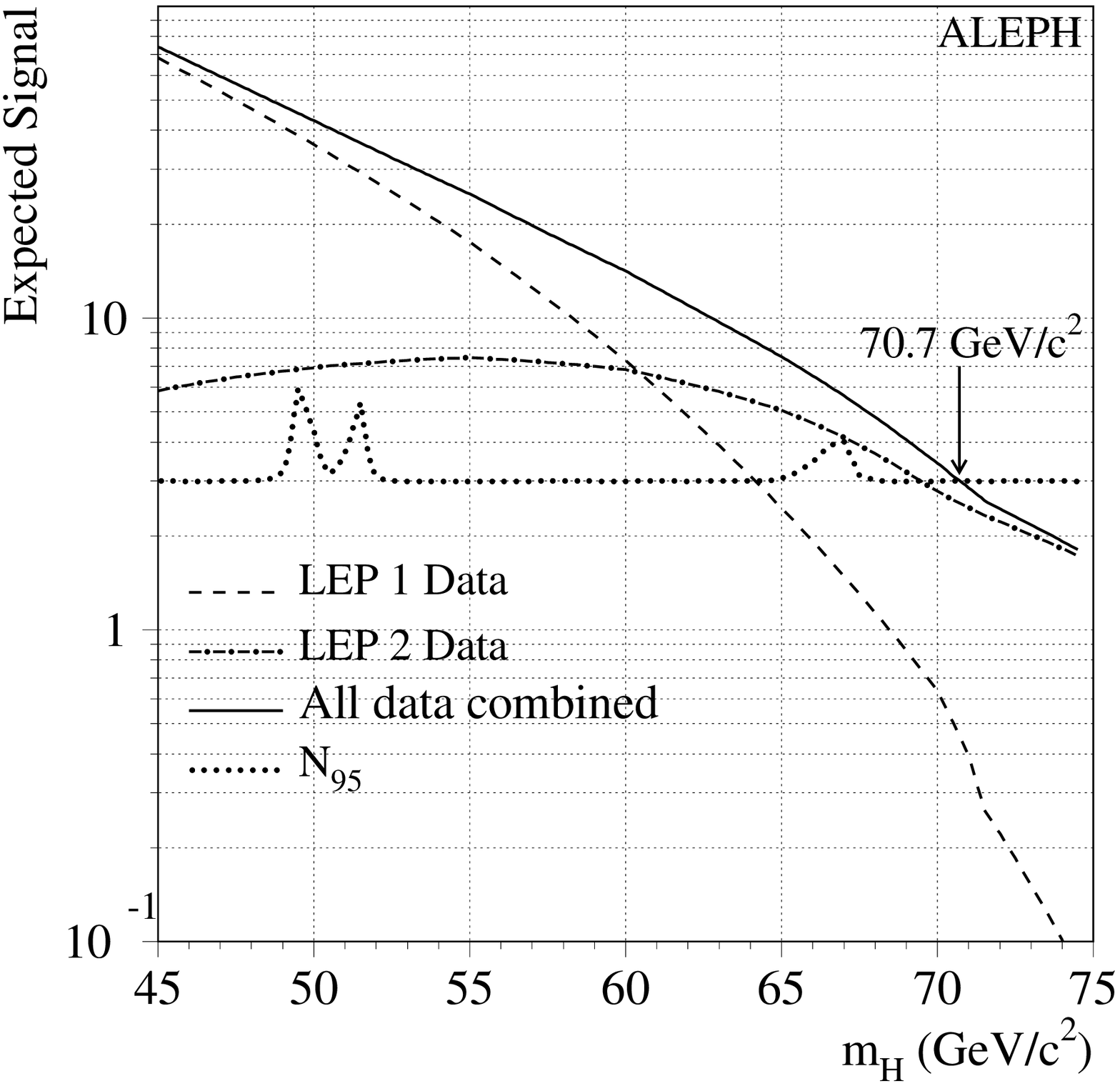}
  \end{center}
\vspace*{-0.8cm}
  \caption[]{\label{fig:aleph} 
             ALEPH MSM Higgs boson mass limit
             including data up to 172~GeV.}
\vspace*{0.5cm}
  \begin{center}
    \includegraphics[width=\textwidth]{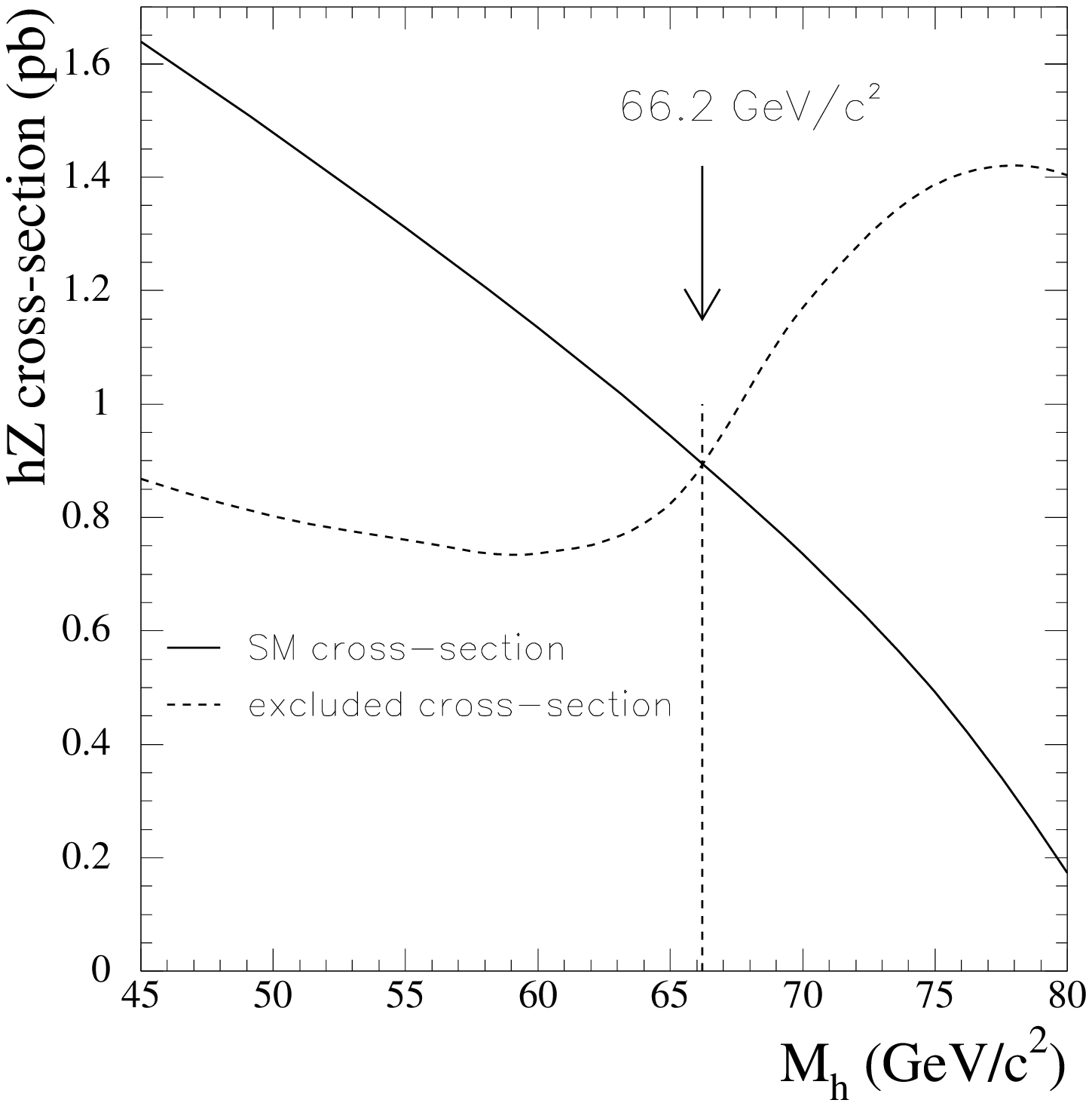}
  \end{center}
\vspace*{-0.8cm}
  \caption[]{\label{fig:delphi}
             DELPHI MSM Higgs boson mass limit
             including data up to 172~GeV.}
\end{minipage}
\hfill
\begin{minipage}{0.48\textwidth}
\vspace*{0.9cm}
  \begin{center}
    \includegraphics[width=\textwidth]{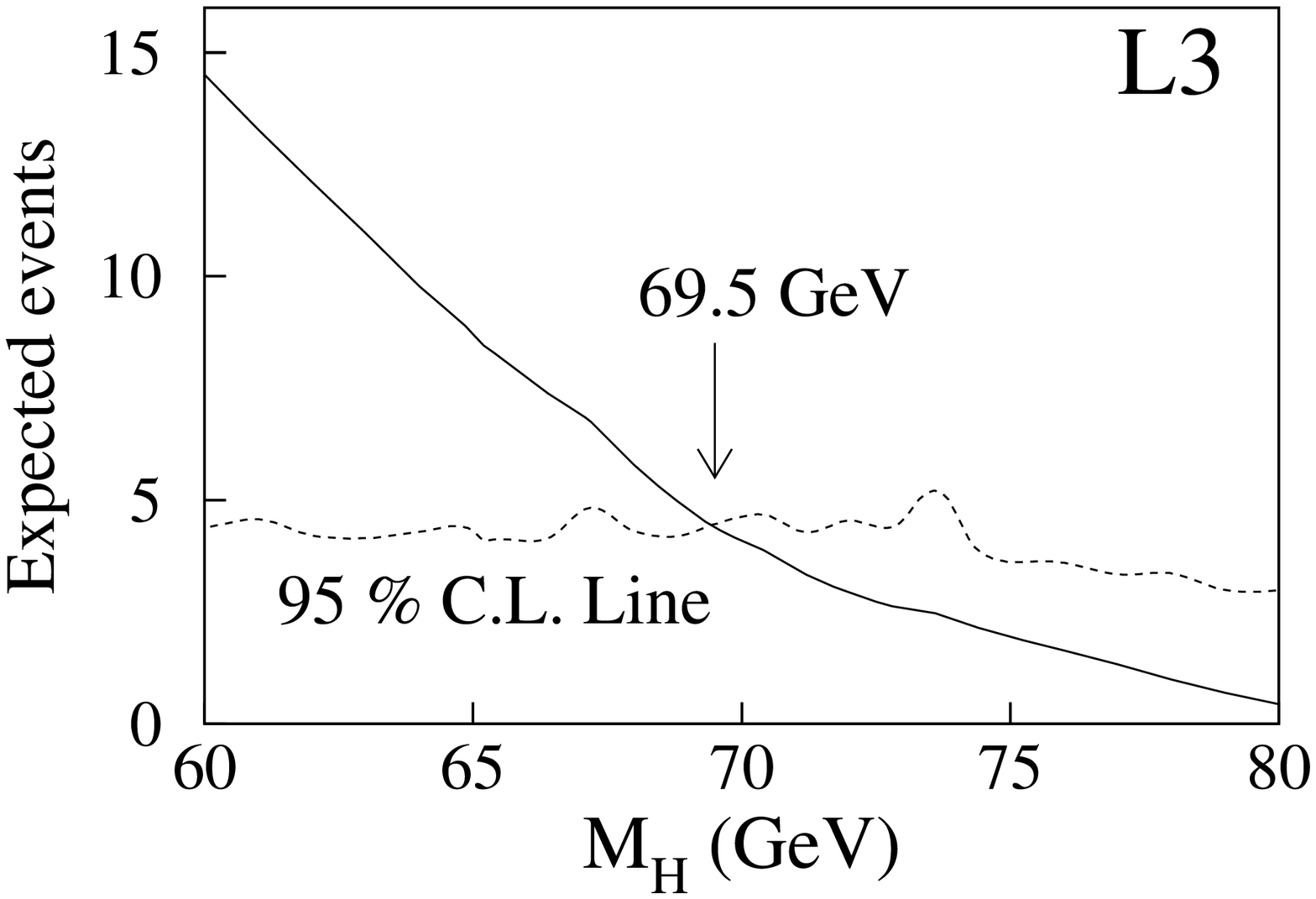}
  \end{center}
\vspace*{-0.1cm}
  \caption[]{\label{fig:l3}
             L3 MSM Higgs boson mass limit
             including data up to 172~GeV.}
\vspace*{1.0cm}
  \begin{center}
    \includegraphics[width=\textwidth]{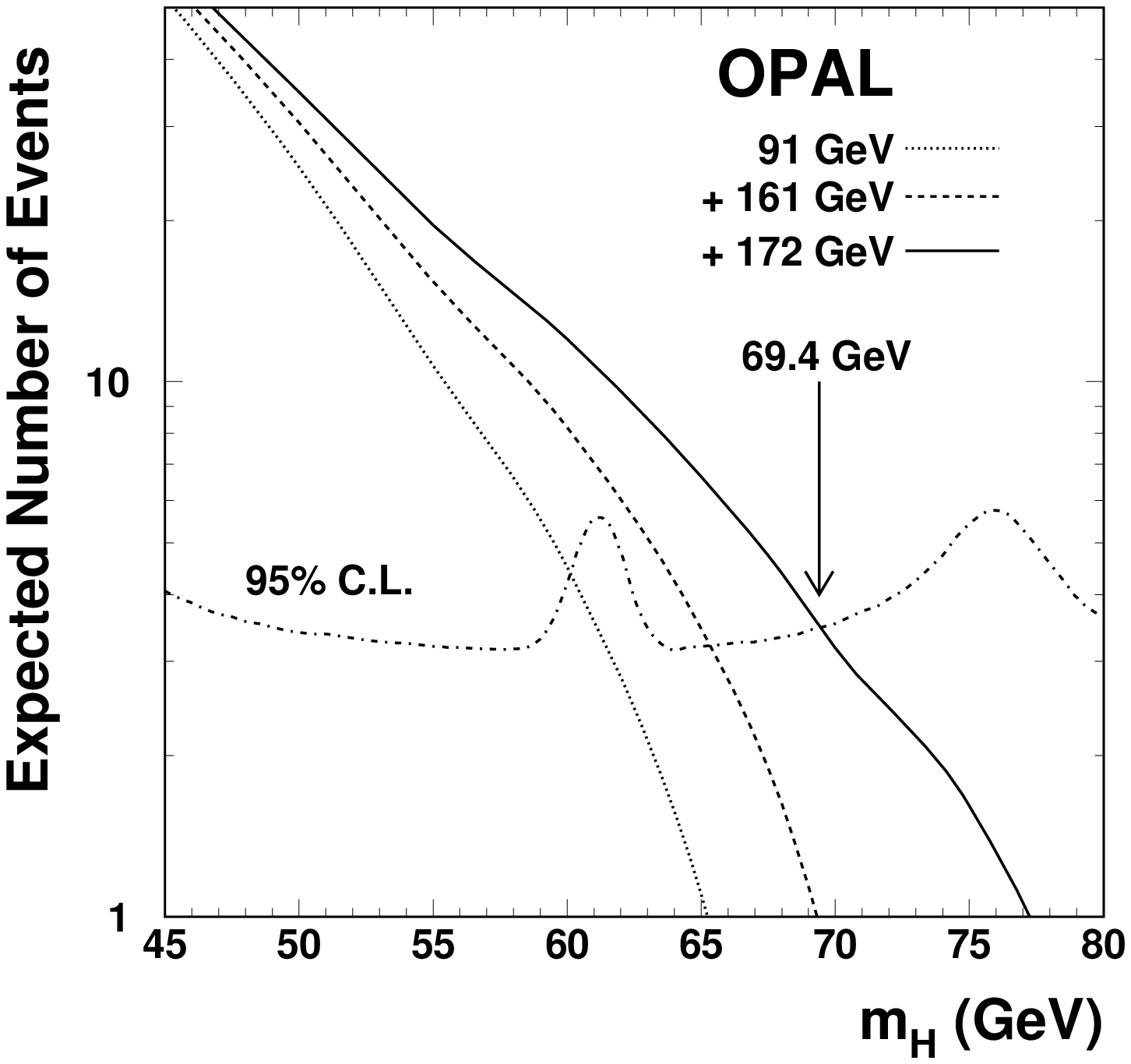}
  \end{center}
\vspace*{-0.5cm}
  \caption[]{\label{fig:opal}
             OPAL MSM Higgs boson mass limit
             including data up to 172~GeV.}
\end{minipage}

\end{figure}

Lower limits on the Higgs boson mass are derived. In the absence of
candidates the 95\% CL limit is set where 3.0 signal events are expected.
Candidates increase the number of expected signal events required according
to their mass resolution.
An overview of the mass limits is given in Table~\ref{tab:limits} and 
details for each experiment are shown in
Figs.~\ref{fig:aleph} 
to~\ref{fig:opal} (from~\cite{aleph,delphi,l3,opal}).

Very preliminary results of the 1997 data-taking at 183~GeV are summarized 
in Table~\ref{tab:1997} (from~\cite{lepc97}). 
The larger part of the data collected in 1997 is analysed using mostly the
analyses tuned for lower energies. Hence, sensitivities are expected 
to increase by optimization for the new data. At a later stage the
combination of the data from the LEP experiments will significantly
increase the sensitivity.

\begin{table}[htp]
\begin{center}
\vskip 0.5cm
\begin{tabular}{|c|cc|cc|cc|}
\hline
\rule{0pt}{12pt}
Final state  
& \multicolumn{2}{|c|}{Background} 
& \multicolumn{2}{|c|}{Efficiency(\%)} 
& \multicolumn{2}{|c|}{$N_{\mathrm{exp}}$}\\ \hline 
\multicolumn{7}{|c|}{ALEPH} \\ \hline
$Hqq$      & 0.17 & 0.23  & 21.1 & 21.9  & 0.24 & 1.12 \\ 
$H\nu\nu$  & 0.06 & 0.09  & 26.3 & 42.9  & 0.11 & 0.70 \\
$H(ee,\mu\mu)$ & 0.06 & 0.11  & 64.2 & 74.8  & 0.08 & 0.40 \\
$H\tau\tau$& 0.02 & 0.02  & 18.8 & 20.4  & 0.01 & 0.05 \\
$\tau\tau qq$& 0.05 & 0.03  & 17.4 & 17.4  & 0.02 & 0.07 \\ \hline
Total       & 0.36 & 0.48  &      &       & 0.46 & 2.34 \\ \hline\hline
\multicolumn{7}{|c|}{DELPHI} \\ \hline
   $Hqq$   & 0.30 & 0.50 & 21.6 & 23.6 & 0.25 & 1.21 \\  
   $H\nu\nu$& 0.65 & 0.61 & 36.3 & 42.8 & 0.12 & 0.63 \\ 
   $Hee$   & 0.13 & 0.20 & 41.7 & 37.2 & 0.02 & 0.09 \\
   $H\mu\mu$& 0.04 & 0.13 & 69.0 & 69.8 & 0.04 & 0.17 \\ 
   $qq\tau\tau$& 0.31 & 0.22 & 22.9 & 24.4 & 0.01 & 0.06 \\ 
   $\tau\tau qq$& 0.32 & 0.91 & 22.1 & 24.4 & 0.02 & 0.10 \\\hline
   Total   & 1.74 & 2.50 &      &      & 0.46 & 2.26 \\\hline
\multicolumn{7}{|c|}{L3} \\ \hline
$qqqq$       & 0.77  &  3.68 & 28.1 & 38.5 & 0.37  & 1.87  \\  
$qq\nu\nu$   & 0.40  &  1.46 & 46.0 & 69.4 & 0.17  & 0.97  \\ 
$qqee$       & 0.03  &  0.18 & 45.5 & 65.8 & 0.03  & 0.15 \\
$qq\mu\mu$   & 0.04  &  0.15 & 34.4 & 48.3 & 0.02  & 0.11 \\ 
$qq\tau\tau$ & 0.008 &  0.23 & 13.5 & 34.9 & 0.01 & 0.08  \\ 
$\tau\tau qq$& 0.0   &  0.25 &  0.0 & 17.7 & 0.0   & 0.07 \\\hline
   Total     & 1.25  &  5.96 &   &       & 0.60  & 3.26\\\hline
\multicolumn{7}{|c|}{OPAL} \\ \hline
   $bbqq$     & 0.75 & 0.88 & 30.8 & 28.2 & 0.35 & 1.29 \\ 
   $qq\nu\nu$ & 0.90 & 0.55 & 38.1 & 41.3 & 0.16 & 0.66 \\ 
   $Hee$      & 0.06 & 0.08 & 52.6 & 65.3 & 0.04 & 0.17 \\ 
   $H\mu\mu$  & 0.04 & 0.06 & 67.8 & 70.6 & 0.04 & 0.18 \\ 
   $qq\tau\tau$& 0.10 & 0.41 & 17.3 & 21.7 & 0.01 & 0.05 \\ 
   $\tau\tau qq$& 0.06 & 0.18 & 17.0 & 19.3 & 0.02 & 0.08 \\\hline
   Total      & 1.91 & 2.16 &      &      & 0.62 & 2.43 \\\hline
\end{tabular}
\end{center}
\caption{\label{tab:selection}
Expected background,
signal efficiency, and the expected numbers of 
signal events for a Higgs boson mass of 70~GeV. 
In each column, the entries on the left are for 161~GeV and those
on the right for 170 to 172~GeV. 
}
\vspace*{-1.cm}
\end{table}

\begin{table}
\begin{minipage}{0.7\textwidth}
\begin{tabular} {|c|c|c|c|} \hline
Experiment & Channel  &$\sqrt{s}$ (GeV) &  Mass (GeV)        \\ \hline
           &          &                 &                    \\
DELPHI     & $Hqq$       &   172           &   $58.7\pm3.5$     \\
           & $H\nu\nu$       &   161           &   $64.6^{+5}_{-3}$  \\
             &          &                 &                    \\ 
OPAL       & $Hqq$       &   172           &   $75.6\pm3.0$     \\
           & $H\nu\nu$   &   161           &   $39.3\pm4.9$ \\\hline
\end{tabular}
\end{minipage}
\hfill
\begin{minipage}{0.29\textwidth}
\vspace*{-0.5cm}
\caption{\label{tab:delopal} 
         Candidate events from DELPHI and OPAL. Their number is in
         agreement with the background expectation and no peak in the 
         reconstructed invariant mass distribution is observed.} 
\end{minipage}
\end{table}

\begin{table}
\begin{minipage}{0.7\textwidth}
\begin{tabular} {|c|c|c|c|c|} \hline
Experiment & Limit(GeV)  & Limit(GeV) & Background & Candidate \\
data set   & 161/172      & \& LEP1     & 161/172    & 161/172   \\  \hline
           &              &             &            &           \\ 
ALEPH      & 69.6         &  70.7         & 0.84       & 0         \\
           &              &               &            &           \\
DELPHI     & 66.2         &  ---          & 4.24       & 2         \\
           &              &               &            &           \\
L3         & 69.3         &  69.5         & 38.1       & 33        \\
           &              &               &            &           \\
OPAL       & 68.9         &  69.4         &  4.07      & 2 \\\hline
\end{tabular}
\end{minipage}
\hfill
\begin{minipage}{0.25\textwidth}
\caption{\label{tab:limits} 
         Individual Higgs boson mass limits at 95\% CL 
         for 161 to 172~GeV data, and 
         in combination with LEP1 data. The numbers of background and 
         candidate events are given.} 
\end{minipage}
\end{table}

\mbox{ }
\vspace*{-4.5cm}

\begin{table}[H]
\vspace*{0.5cm}
\begin{minipage}{0.45\textwidth}
\begin{tabular} {|c|cc|c|}\hline
Experiment & \multicolumn{2}{|c|}{Limit (GeV)}& {$\cal L$} (pb$^{-1}$) \\\hline
           &            &               &      \\
ALEPH      & 83         &    88.6       & 55   \\
           &            &               &      \\
DELPHI     & 85.8       &    83.6       & 53.5 \\
           &            &               &      \\
L3         & 81.3       &    82.2       & 36   \\
           &            &               &      \\
OPAL       & $-$         &    82         & 39   \\ \hline
\end{tabular}
\end{minipage}
\hfill
\begin{minipage}{0.54\textwidth}
\vspace*{-0.8cm}
\caption{\label{tab:1997}
Preliminary expected (left) and observed (right) MSM Higgs boson mass limits at 95\%~CL
for the luminosity $\cal L$ of analysed 183~GeV data.
ALEPH, DELPHI and L3 have also released their preliminary expected mass 
limits. Higher (lower) observed limits than expected limits correspond to 
slightly less (more) data events observed than expected from the background.}
\end{minipage}
\vspace*{-0.5cm}
\end{table}

\clearpage
\section{\label{sec:msmcombi} Combined MSM Higgs Boson Results}

Much progress has been made recently in combining the LEP 
results~\cite{lepwg}.
Different statistical methods of ALEPH~\cite{jadib}, DELPHI~\cite{alex},
L3~\cite{l3method}, and OPAL~\cite{bock}
are used and compared in order to derive a 95\% CL lower
limit on the MSM Higgs mass. The different statistical methods satisfy 
the following criteria:

\begin{itemize}
\item  The limit should be at least at the 95\% CL and come close to 
       the desired 5\% false exclusion rate of the Higgs boson hypothesis.

\item  The order of the combination of different channels 
       should not change the final limit.

\item  The expected limit from combining two channels should exceed the
limit from any single channel.

\item  Candidate events which are incompatible with the Higgs boson 
       hypothesis should not change the mass limit.
\end{itemize}

The fact that a background fluctuation should not give a stronger limit on
the Higgs boson hypothesis is taken into account by DELPHI, L3 and OPAL
using the modified frequentist definition of the confidence level~\cite{PDG}:
$$1-CL=P(X_{\mathrm{s+b}}\leq X_{\mathrm{obs}}) /
       P(X_{\mathrm b}\leq X_{\mathrm{obs}}),$$
while ALEPH uses a more conservative definition 
without background subtraction:
$$1-CL=P(X_{\mathrm s}\leq X_{\mathrm{obs}}),$$
where $X_{\mathrm s},X_{\mathrm b},$ and $ X_{\mathrm{obs}}$ 
stand for signal,
background, and observed distribution functions to separate signal
and background.
Both definitions give better confidence levels than the claimed ones, 
however, the former definition tends to be closer to the claimed 
confidence level, especially for a large number of background events.

The following methods are used by the four LEP experiments to set a limit:

\begin{itemize}
\item  ALEPH: The distribution function is defined as
$$X=\prod_{i=1}^{n}c_{i}^{a_{i}}(m_{H}),$$
where $c_{i}$ are the confidence levels for $i=1,...,n$ different channels,
and $a_{i}$ are weight factors to ensure that an additional channel never
degrades the confidence level.
In this way, first experiments derive their own limits and, in a second
step, the results are combined using a simple analytic function.

\item  DELPHI: The method uses a likelihood ratio which can be readily
derived using Poisson statistics:

$$Q(m_{H})=\prod_{i=1}^{n}\mathrm{e}^{-(s_{i}+b_{i})}(s_{i}+b_{i})^{k_{i}}/%
\prod_{i=1}^{n}\mathrm{e}^{-b_{i}}b_{i}^{k_{i}},$$

where $s_{i}$ and $b_{i}$ are the expected signal and background events, and 
$k_{i}$ is the number of observed events for $i=1,...,n$ different channels.
More generally, taking the probability distributions $S_{i}(m_{H},m_{ij})$ and 
$B_{i}(m_{ij})$ of the invariant masses into account, the formula becomes

$$Q(m_{H})=\mathrm{e}^{-s_{\mathrm{tot}}(m_{H})}\prod_{i=1}^{n}%
\prod_{j=1}^{k_{i}} (1 + \frac{ s_{i}(m_{H})S_{i}(m_{H},m_{ij}) }
                         { b_{i}B_{i}(m_{ij}) }) .$$

Weights are defined for each candidate $w_{k}=\ln (1+s_{k}S_{k}/b_{k}B_{k})
$ where $\ln Q=-s_{\mathrm{tot}}+X$ $\ $with $X=\sum_{k=1}^{N}n_{k}w_{k}$ and $N$
the total number of candidates. With this distribution function,
the confidence level is defined as

$$1-CL=P(\sum_{k=1}^{n}n_{k}^{\mathrm{s+b}}w_{k}\leq
\sum_{k=1}^{n}n_{k}^{\mathrm{obs}}w_{k}) ~/~
P(\sum_{k=1}^{n}n_{k}^{\mathrm b}w_{k}\leq
\sum_{k=1}^{n}n_{k}^{\mathrm{obs}}w_k),$$

where the distribution of $n_{k}^{\rm s+b}$ and $n_{k}^{\rm b}$ are 
determined with MC simulation using Poisson statistics.

\item  L3: As for DELPHI, a likelihood function is defined
for the {\it signal + background} hypothesis:
$$L(s,b) = 
e^{-(s+b)}\prod^{k}_{j=1}\frac{(s\cdot f_j + b\cdot g_j)^{n_j}}{n_j!}.$$
The index $j$ runs over all $k$ channels;
$f_j$, $g_j$ are the fractions of the total signal $s$ and  
total background $b$, respectively, and  
$n_j$ is the number of candidate events which fall into channel $j$.

The Bayesian confidence interval is used as distribution function.
This requires an additional integration over the signal events:
$$
 X = \frac{\int_{\mu ^{H}}^{\infty} L(s,b) ds}
{\int_{0}^{\infty} L(s,b) ds} ,$$
where $\mu ^{H}$ is the number of expected signal events, depending on the 
Higgs boson mass hypothesis.
In order to come closer to the desired false exclusion rate,
$1-CL$ is defined as in the DELPHI case using MC simulations for 
signal-and-background, and signal-only hypotheses.

\item  OPAL: A weight is assigned to each candidate event using a weight
function $F(m_{i}),$ where $m_{i}$ is the mass of the $i$th candidate.
The weight function is based on the difference between a Higgs mass
hypothesis and the actual mass distribution of candidate $i$. In order to
account for different signal and background ratios in channel $k$, the
following weight function is defined:

$$F_{k}(m)=
K(C+b_{k}s_{\mathrm{tot}}/D_{k}^{\max }s_{k})^{-1}D_{k}(m)/D_{k}^{\max },$$

where $K$ is chosen to set the largest value of $F_{k}$ to unity, and $D$ is
the mass distribution. The result is almost independent of the constant $C$.
The distribution function is defined as
$$X=\sum_{k=1}^{n} F_k(m),$$
where $n$ is the number of candidates in all channels.
\end{itemize}

\begin{figure}[htp]
\begin{center}
\includegraphics[width=8.4cm]{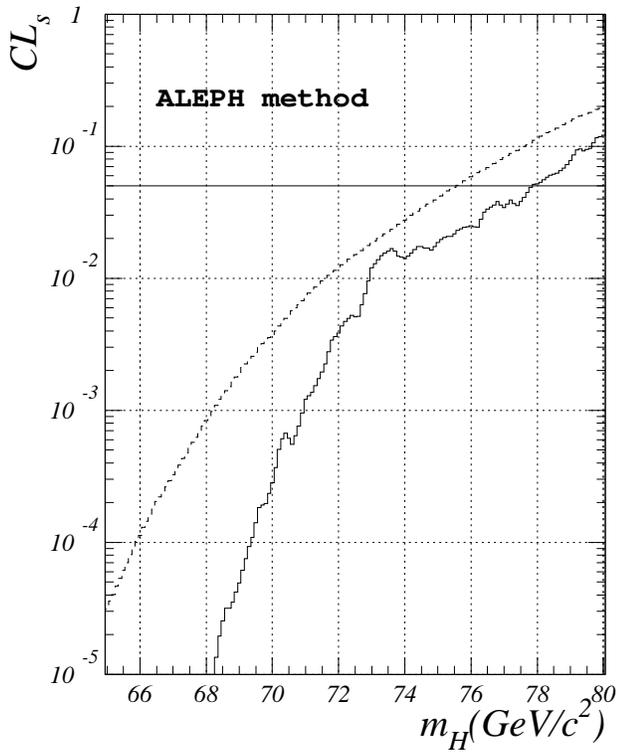} 
\includegraphics[width=8.4cm]{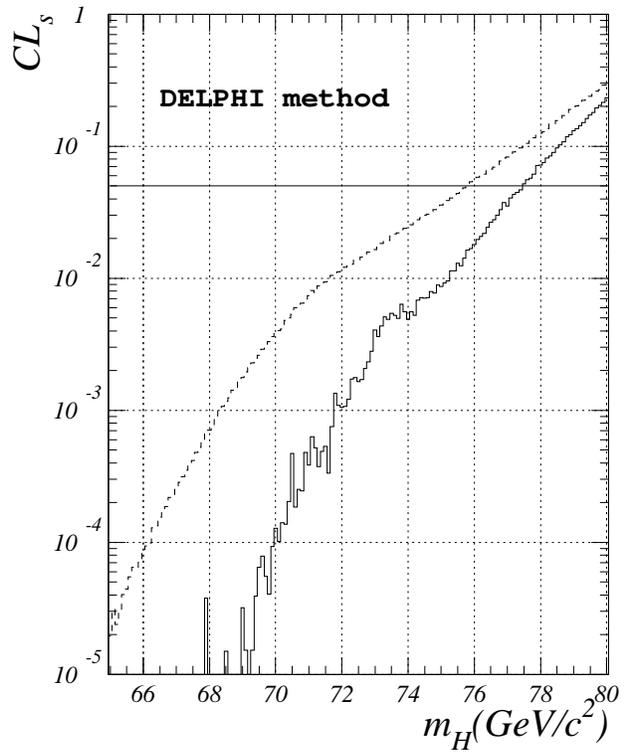} 
\includegraphics[width=8.4cm]{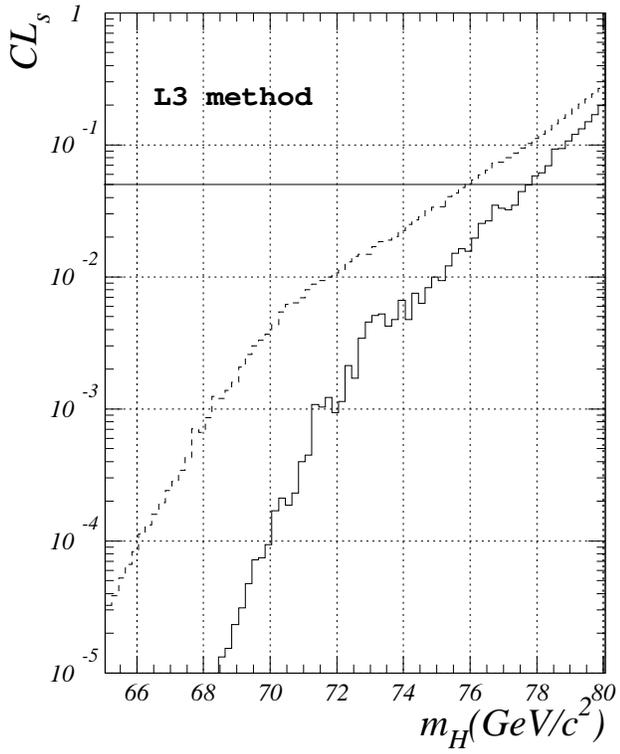} 
\includegraphics[width=8.4cm]{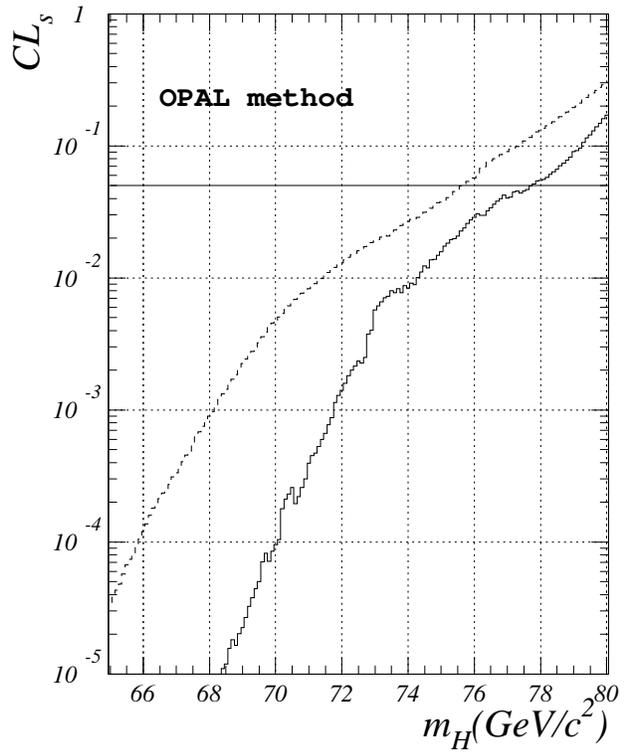} 
\caption[]{Expected (dashed line) and observed (full line) 
confidence levels, $1-CL \equiv CL_{\mathrm s}$,
as a function of the hypothetical Higgs boson mass,
obtained from combining the results of the LEP experiments 
using the four statistical methods.
The intersections of the curves with the 5\% horizontal line define the 
expected and observed 95\% CL lower bounds for the mass of the MSM
Higgs boson.
\label{fig:lep_cls}}
\end{center}
\end{figure}

Figure~\ref{fig:lep_cls} (from~\cite{lepwg}) shows the combined Higgs 
boson mass limits using the four statistical methods.
The small differences in the results are in part attributed to the 
different statistical ways of treating with candidates close to the 
mass limits.
Table~\ref{tab:com_limits}\footnote{Small differences in comparison with
Table~\protect{\ref{tab:limits}} exist because of the updated statistical 
methods.} 
(from~\cite{lepwg}) compares the expected 
and observed mass limits for the four statistical methods. The observed 
mass limits are about 2~GeV larger, since the number of
observed candidate events is slightly below the number of expected
background events. For the combined limit, the background-only confidence
level is between 26\% and 35\% depending on the statistical method, and 
thus the data are consistent with the background.

\begin{table}[htp]
\vspace*{-0.2cm}
\begin{center}
\begin{tabular}{|c|cc|cc|cc|cc|}
\hline
\rule{0pt}{12pt}
Experiment  & \multicolumn{8}{|c|}{Statistical method} \\
            & \multicolumn{2}{|c|}{ALEPH} 
            & \multicolumn{2}{|c|}{DELPHI} 
            & \multicolumn{2}{|c|}{L3}   
            & \multicolumn{2}{|c|}{OPAL}                          \\\hline
 ALEPH      & 68.5 & 69.6& 68.5 & 69.6 & 68.8 & 69.6 & 68.6 &69.6 \\
 DELPHI     & -    & -   & 65.4 & 65.9 & 65.3 & 65.9 & 65.1 &65.5 \\
 L3         & -    & -   & 66.1 & 69.4 & 65.7 & 69.3 & 65.0 &68.2 \\
 OPAL       & -    & -   & 65.9 & 69.0 & 65.6 & 68.6 & 65.3 &68.9 \\\hline
 LEP        & 75.7 & 77.9& 75.8 & 77.5 & 76.0 & 77.8 & 75.6 &77.7 \\\hline
 \end{tabular}
\end{center}
\vspace*{-0.2cm}
\caption{\label{tab:com_limits}
{Expected (left) and observed (right) 95\% CL mass 
limits (in GeV) for the individual
experiments and for LEP combined using, in each case, 
the four statistical methods.
The entries in the first column pertaining to the DELPHI, 
L3 and OPAL experiments are empty  since the ALEPH method 
does not recalculate the individual limits.}
}
\end{table}

The interest of electroweak working groups in the direct
limits resulted in the following proposal on 
how to combine the results~\cite{lepwg}.
First, note that the information given is more 
detailed than for a Higgs boson mass limit at the 95\% CL.
Furthermore,
the confidence levels are either less than or equal to
the signal probability. If one wishes to combine these Higgs boson
confidence levels with line-shape data using a $\chi^2$ method, the
interpretation of the confidence-level as a probability and the conversion
of the confidence-level values would lead to a lower bound on the 
$\Delta \chi^2$.
If one wishes to set upper Higgs boson mass limits in combination
with other data using Bayesian methods, the lower $\Delta \chi^2$ bound would 
not be conservative.
Consequently, the given confidence levels are not suitable as a basis
for combining the direct limits with other results.

\begin{figure}[htbp]
\begin{minipage}{0.6\textwidth}
\vspace*{-1cm}
  \begin{center}
    \includegraphics[width=\textwidth]{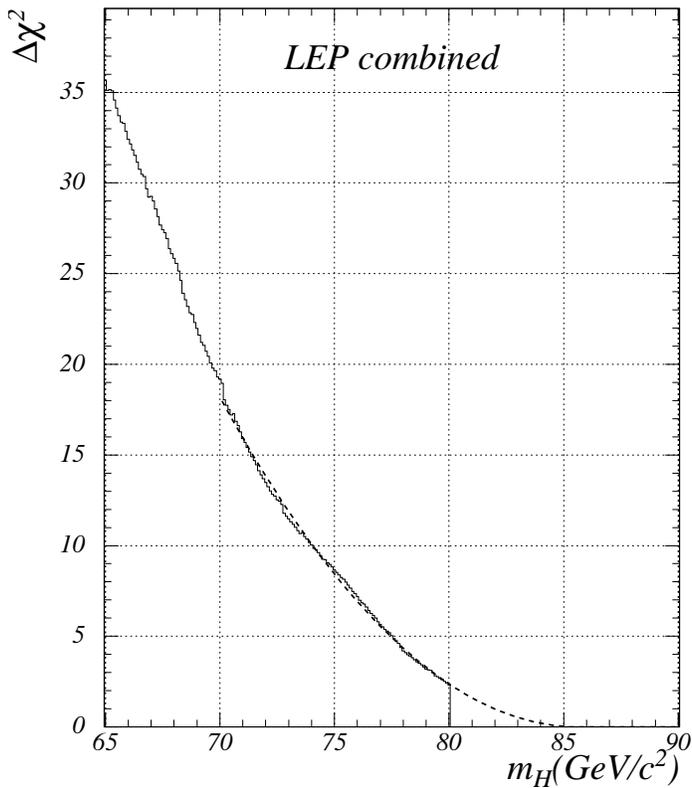}
  \end{center}
\end{minipage}
\hfill
\begin{minipage}{0.39\textwidth}
  \caption[]{\label{fig:chi2}
             $\Delta \chi^2$ as a function of the Higgs boson mass
             from direct searches (solid line).
             The extrapolation to $m_{H} > 80$~GeV is obtained by a
             parabolic fit in the domain $70 < m_{H} < 80$~GeV.
             }
\end{minipage}
\vspace*{-1cm}
\end{figure}

A $\Delta \chi^2$ defined from a likelihood ratio could
then directly be combined with indirect results. The likelihood ratio 
$Q(m_H) \equiv L_{s+b} / L_{b}$
corresponding to the signal-and-background and the background-only hypotheses
is given in the DELPHI statistical method.
Figure~\ref{fig:chi2} (from~\cite{lepwg}) shows the resulting 
$$
\Delta \chi^2 =
-2 (\ln L(m_H) - \ln L(m_H = \infty) ) = 
-2 \ln Q(m_H).
$$
The curve of  $\Delta\chi^2(m_{H})$ obtained in this manner is 
shown up to $m_{ H} = 80$~GeV (solid line). 
The combination of the direct searches from the four experiments was 
not pursued to higher values.
An extrapolation to $m_{H}$ beyond 80~GeV is provided by a parabolic 
fit, performed in the domain $70 < m_{H} < 80$~GeV,
\begin{equation}
   \Delta\chi^2(m_{H}) \approx 0.0743~(m_{H} - 85.7)^2 
   \label{eq:extrap},
\end{equation}
which is shown by the dashed-line curve. 
The extrapolation is a rough estimate, since threshold effects of the Higgs 
boson production cross section are not considered.

This method of deriving a $\Delta \chi^2$ is
identical to the method proposed in~\cite{blondel}
where the likelihood $L_{\rm s+b}$ is used instead of the likelihood
ratio for the special case in which the background does not depend
on the Higgs boson mass hypothesis. However,
for the L3 experiment the likelihood $L_{\rm b}$ is a function
of the Higgs boson mass. 
Furthermore, a Bayesian interpretation
shows that the information gain due to direct searches is given
by the likelihood ratio, when the `a priori' probability for the
signal is small~\cite{jens}.
 
\section{\label{sec:thdm} Two Higgs Doublet Results}

\begin{figure}[ht]
\begin{minipage}{0.48\textwidth}
\vspace*{-0.5cm}
  \begin{center}
    \includegraphics[width=1.1\textwidth]{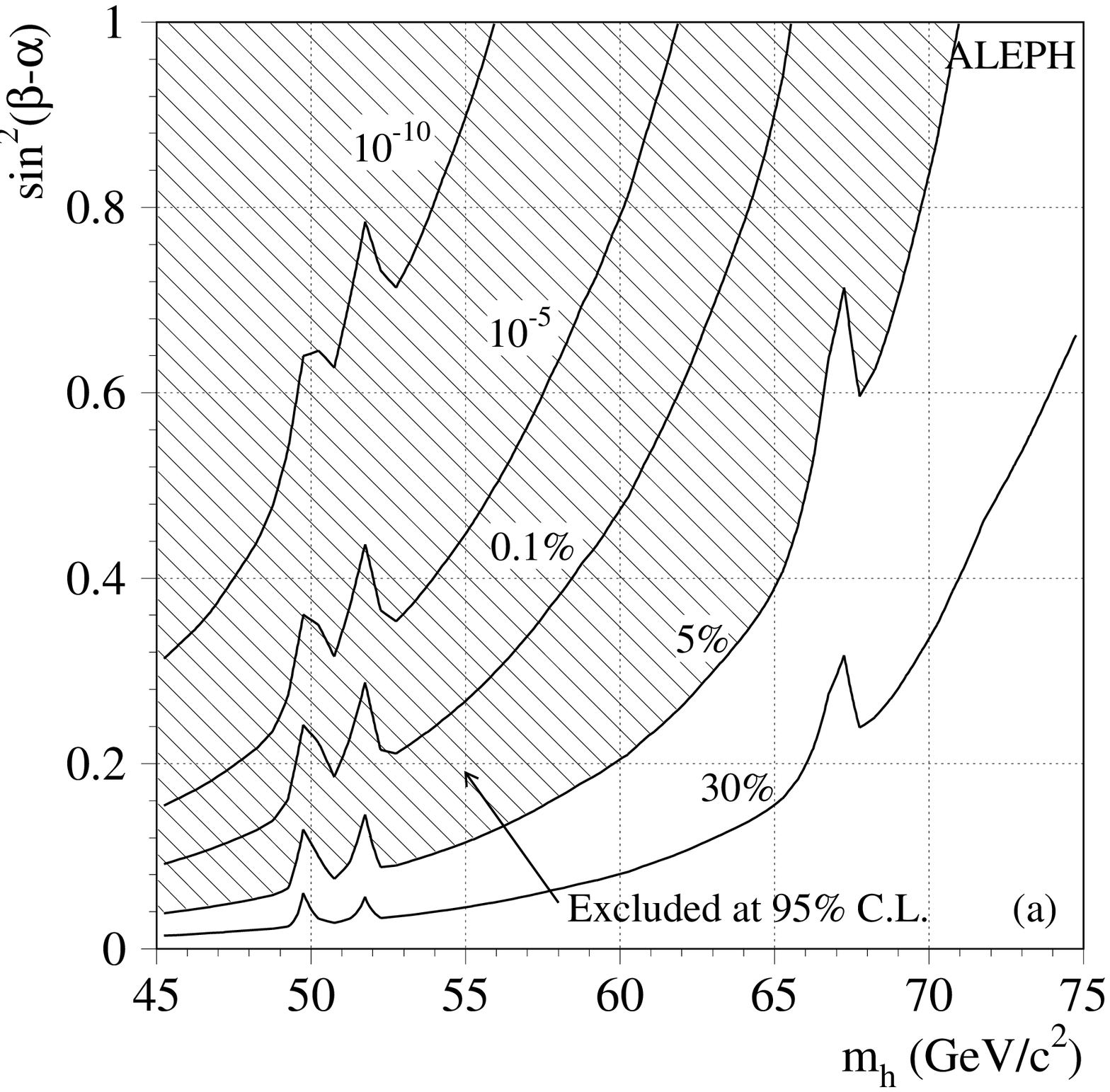}
  \end{center}
\vspace*{-0.9cm}
  \caption{ALEPH $\sin(\beta-\alpha)$ limit.}
  \label{fig:sinba}
\end{minipage}
\hfill
\begin{minipage}{0.48\textwidth}
\vspace*{0.5cm}
  \begin{center}
    \includegraphics[width=\textwidth]{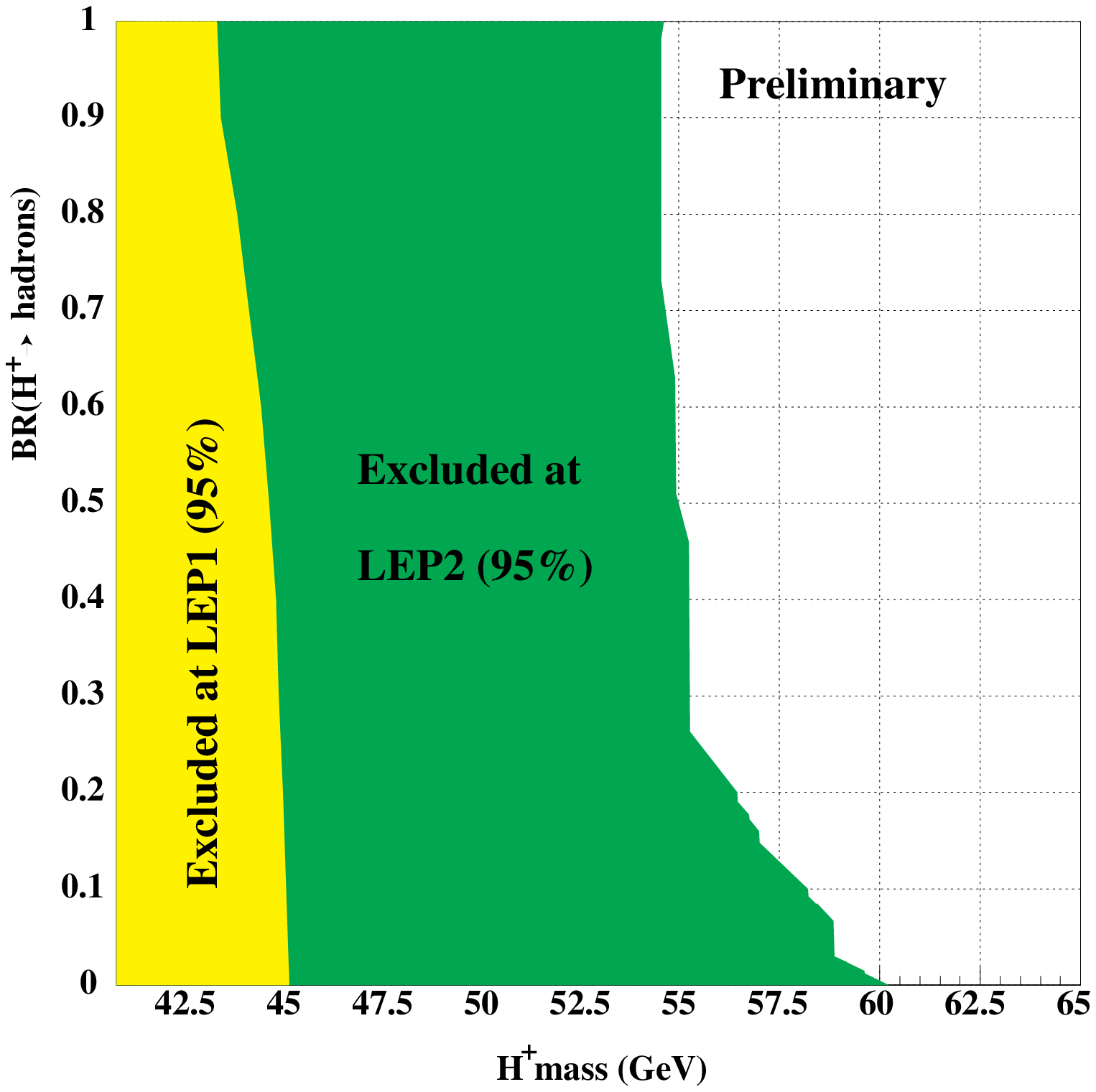}
  \end{center}
\vspace*{-0.7cm}
  \caption{DELPHI charged Higgs boson mass limit.}
  \label{fig:hphm}
\end{minipage}
\end{figure}

In the general framework of the two Higgs doublet model, five physical Higgs
bosons are predicted: two CP-even Higgs bosons, $h$ and $H$, a CP-odd Higgs
boson, $A$, and two charged Higgs bosons, $H^{\pm }$. Searches for these
Higgs bosons are performed in the MSM Higgs boson channels with suppressed
production rates, and for Higgs boson pair-production.
The $\beta $-parameter 
is defined as the ratio of the vacuum expectation values of the two
Higgs doublets and $\alpha $ is the mixing angle of the CP-even Higgs
bosons. The results of the MSM\ Higgs search can be interpreted as limits in
the Higgs mass and $\sin ^{2}(\beta -\alpha )$ parameter plane as shown 
for example in Fig.~\ref{fig:sinba} (from~\cite{alephmssm}). 
These results from ALEPH include the LEP1 data and the effects of their 
three LEP1 candidates are visible.

The charged Higgs boson production and decay processes are:
$$
e^+e^- \rightarrow H^+H^- \rightarrow cscs, cs\tau\nu, ~\mathrm{and}~ \tau\nu\tau\nu.
$$
The resulting signatures are events with four jets, 
two jets, a $\tau$ lepton and missing energy, and two $\tau$ leptons 
with large missing energy. No signal has been observed. The excluded mass 
region is shown in Fig.~\ref{fig:hphm} (from~\cite{delphihphm}) 
as a function of the hadronic decay branching fraction. 
Limits from the four LEP experiments are given in 
Table~\ref{tab:hphm}~\cite{alephhphm,delphihphm,opalhphm,l3hphm}\footnote{Note the indirect limits 
from the decay $b\rightarrow s\gamma$~\protect{\cite{PDG}}.}.

\begin{table}[ht]
\begin{minipage}{0.35\textwidth}
\begin{tabular} {|c|c|c|c|c}\hline
Experiment & Limit (GeV)\\\hline
ALEPH      & 52         \\
DELPHI     & 54.5       \\
L3\footnote{Based on LEP1 data only.}         & 41.0         \\
OPAL       & 52.0       \\\hline
\end{tabular}
\end{minipage}
\hfill
\begin{minipage}{0.64\textwidth}
\caption{\label{tab:hphm}
         Charged Higgs boson mass limits of the four LEP experiments.} 
\end{minipage}
\vspace*{-0.7cm}
\end{table}

\begin{figure}[htbp]
\begin{minipage}{0.48\textwidth}
  \begin{center}
    \includegraphics[width=1.1\textwidth]{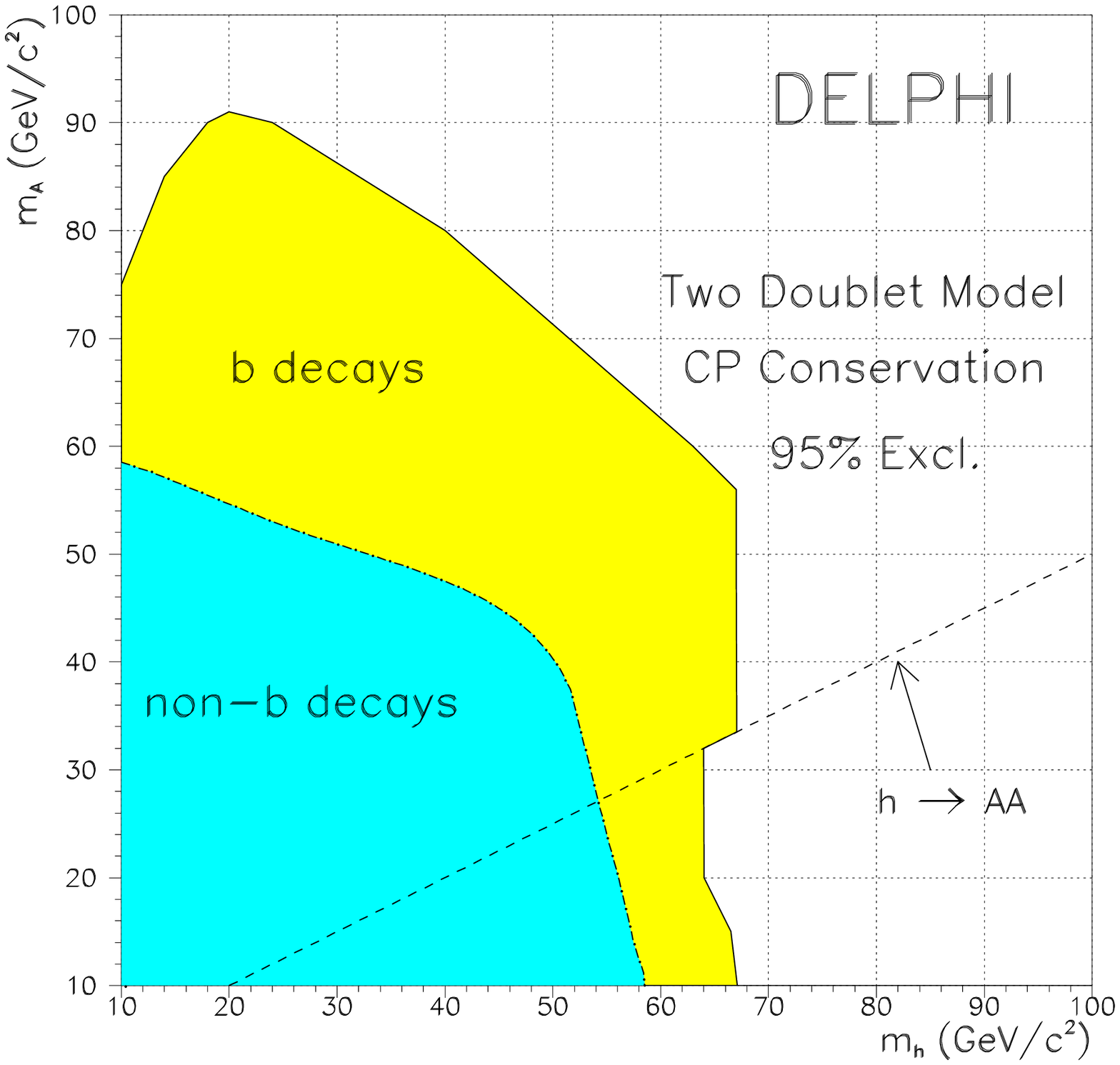}
  \end{center}
\vspace*{-0.5cm}
  \caption{DELPHI $(m_h,m_A)$ limit for CP conservation in the Higgs sector.}
  \label{fig:ha}
\end{minipage}
\hfill
\begin{minipage}{0.48\textwidth}
  \begin{center}
    \includegraphics[width=1.1\textwidth]{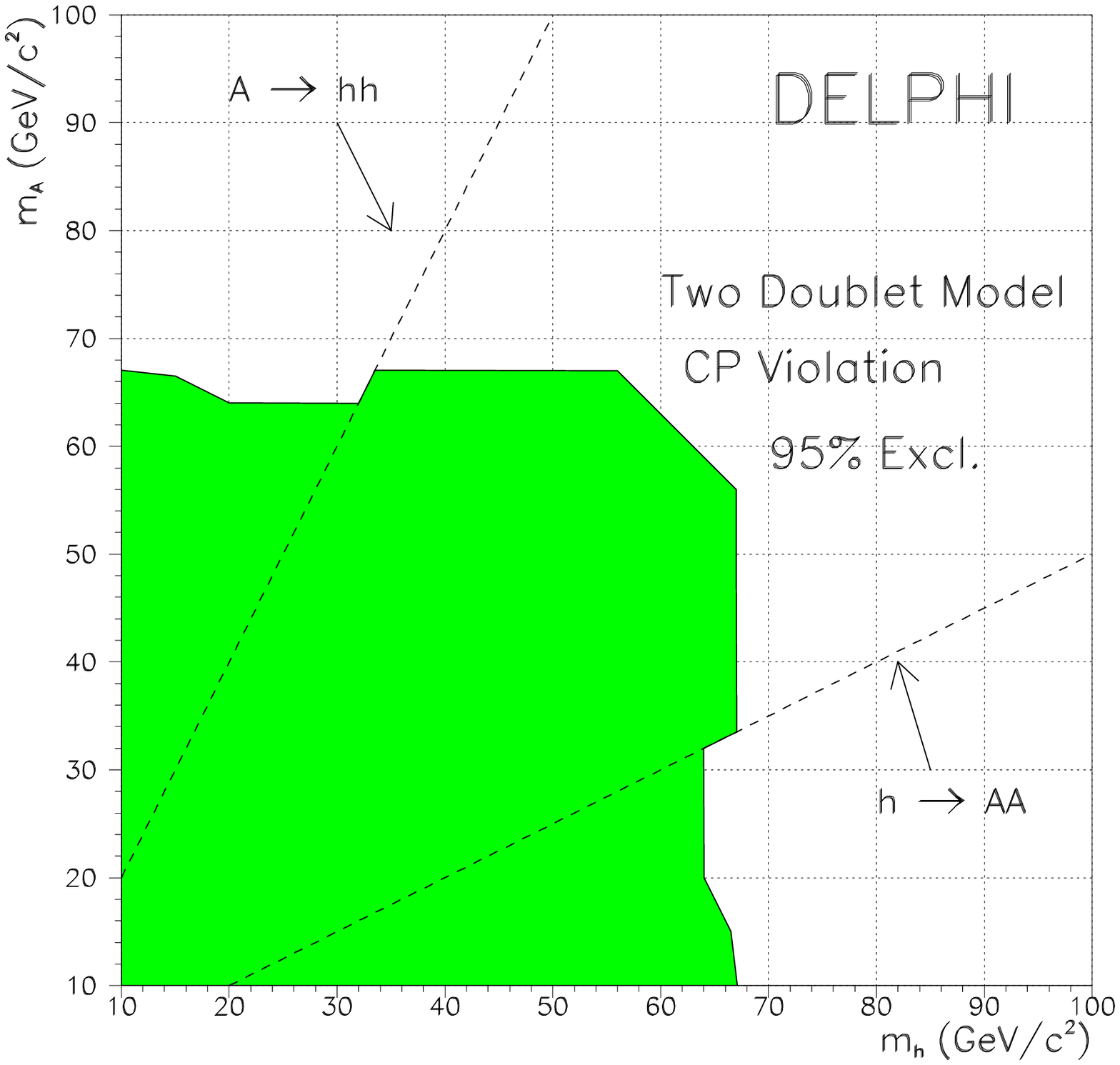}
  \end{center}
\vspace*{-0.5cm}
  \caption{DELPHI $(m_h,m_A)$ limit for CP violation in the Higgs sector.}
  \label{fig:ha2}
\end{minipage}
\end{figure}

The search results for neutral Higgs boson pair-production, 
$e^{+}e^{-}\rightarrow hA$, are presented from DELPHI for the cases in which 
the Higgs bosons decay mainly into $b$-quarks 
(like the MSM Higgs decays) and into
non-b quarks. Since no $b$-tagging can be applied in the latter case, the
detection sensitivity is lower. Note that the decay $h\rightarrow AA$ is
CP-conserving and that this channel leading to six quarks has been
investigated separately. No indication of a signal has been observed and
significantly improved limits compared to LEP1 are given in 
Fig.~\ref{fig:ha} (from~\cite{noncp}). In the more general framework where CP
is not conserved~\cite{jack}, the decay $A\rightarrow hh$ is also possible and 
Fig.~\ref{fig:ha2} (from~\cite{noncp}) shows the resulting limits.

\begin{figure}[htbp]
\begin{minipage}{0.48\textwidth}
  \begin{center}
    \includegraphics[width=1.1\textwidth]{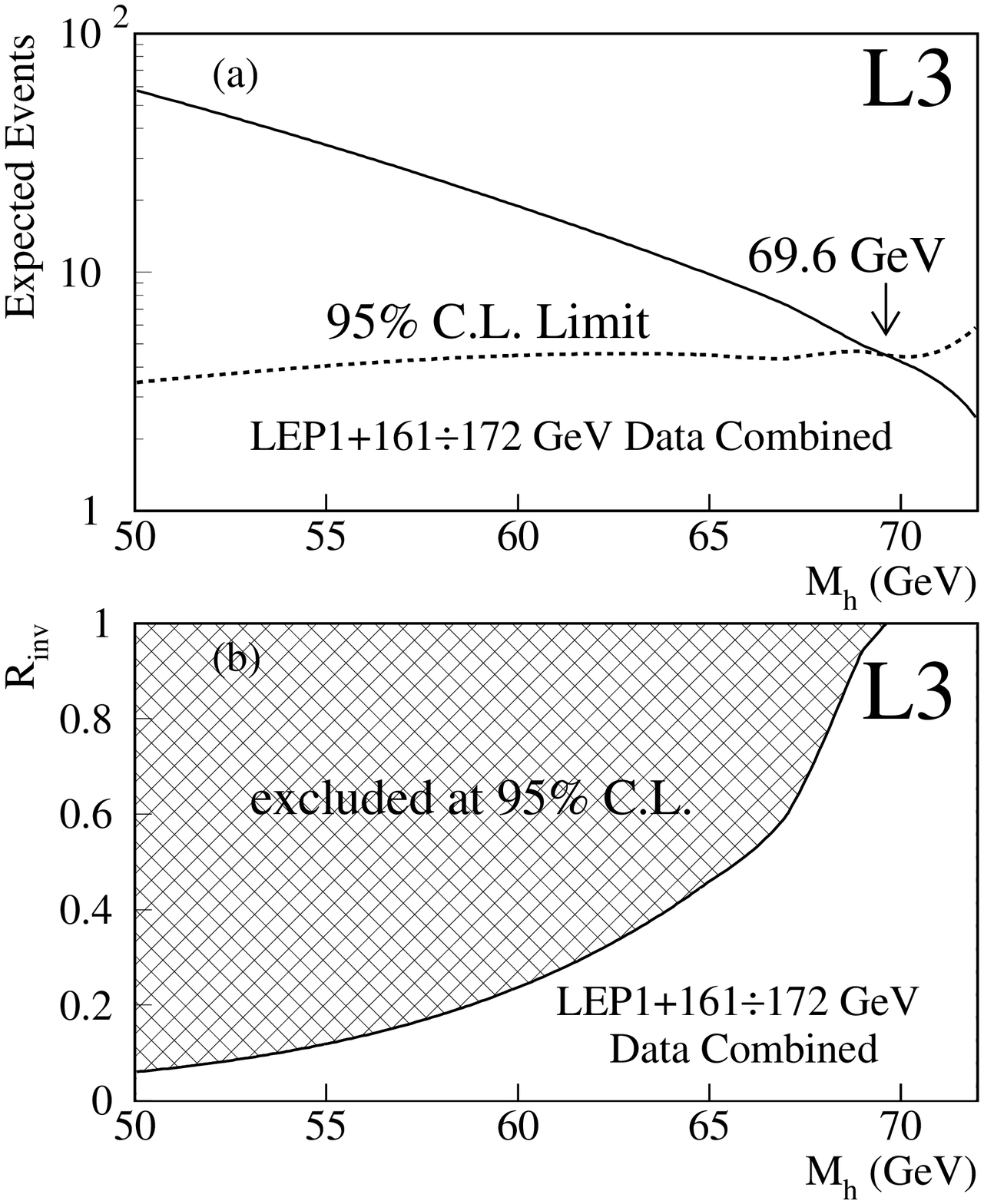}
  \end{center}
\vspace*{-1.2cm}
  \caption{L3 mass limit for invisibly decaying Higgs bosons (a),
           and limits on the production rate as defined in the text (b).}
  \label{fig:invh}
\end{minipage}
\hfill
\begin{minipage}{0.48\textwidth}
  \begin{center}
    \includegraphics[width=1.1\textwidth]{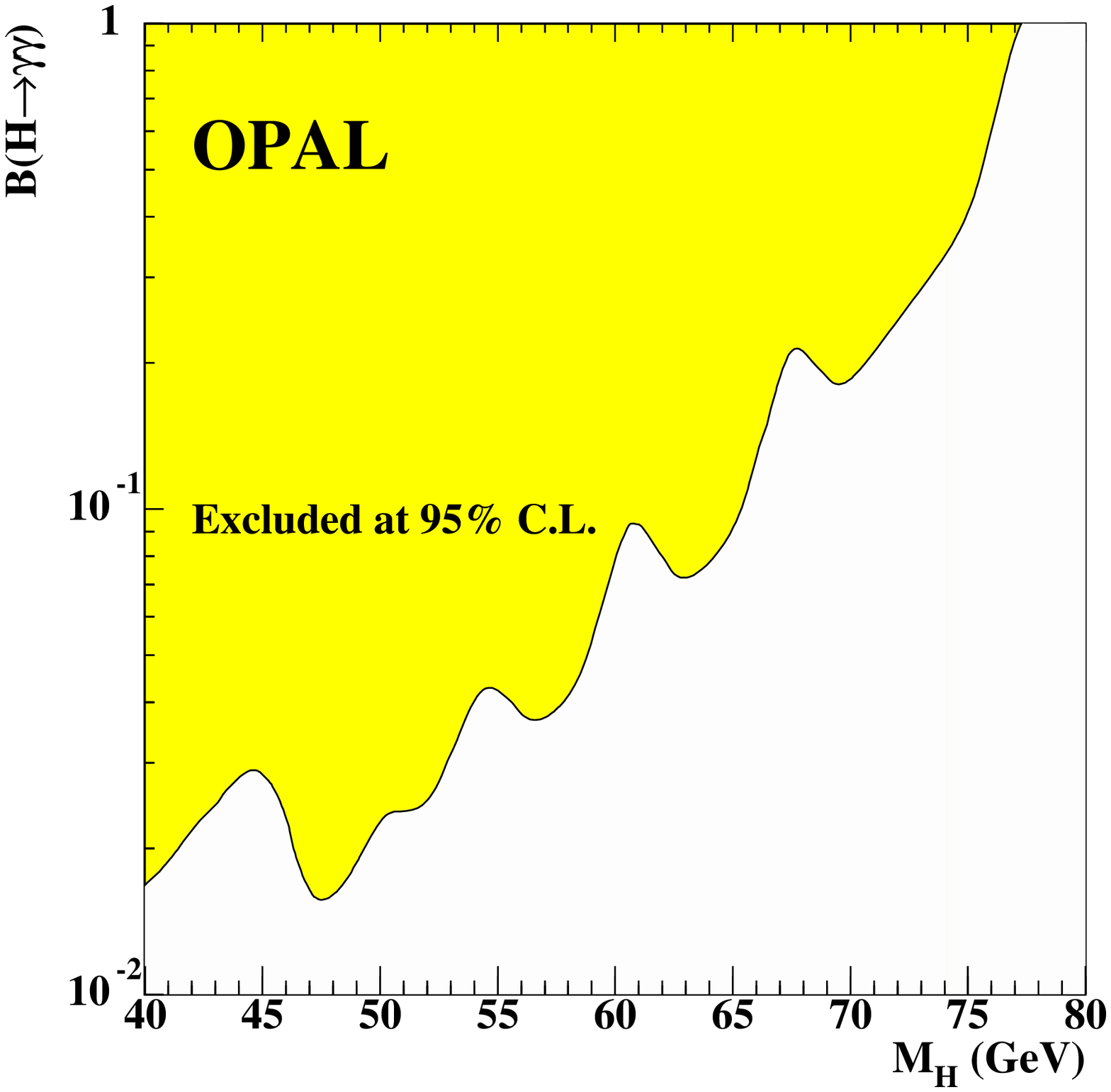}
  \end{center}
  \caption{OPAL branching-fraction limits for Higgs boson decays 
           into a photon pair.}
  \label{fig:hgamma}
\end{minipage}
\end{figure}

In the case of an additional Higgs boson singlet, a massless Higgs particle
exists. It is called the Majoron and does not interact with the 
standard particles.
The massive Higgs bosons could decay into a pair of these massless Higgs
particles and thus their decay would be invisible.
The search signature is very similar to that for the $H\nu\nu$ channel
and the four LEP experiments have set mass limits. For example, 
Fig.~\ref{fig:invh} (from~\cite{l3inv})
gives mass limits under the assumption that the Higgs boson is 
produced with the MSM rate and the decay is completely invisible (plot a),
and plot (b) shows limits on the production ratio 
$R_{\rm inv} \equiv \sigma(Zh)BR(h\ra {\rm invisible}) / \sigma(ZH_{\rm MSM})$.

Based on general searches for a photon pair and missing energy,
limits are also set on the branching fraction 
$BR(H\rightarrow \gamma\gamma)$
as shown for example in Fig.~\ref{fig:hgamma} (from~\cite{opalhgamma}).

\newpage
\section{\label{sec:mssm} MSSM Interpretation}

\begin{figure}[htbp]
\begin{minipage}{0.57\textwidth}
\vspace*{-1.1cm}
  \begin{center}
    \includegraphics[width=\textwidth]{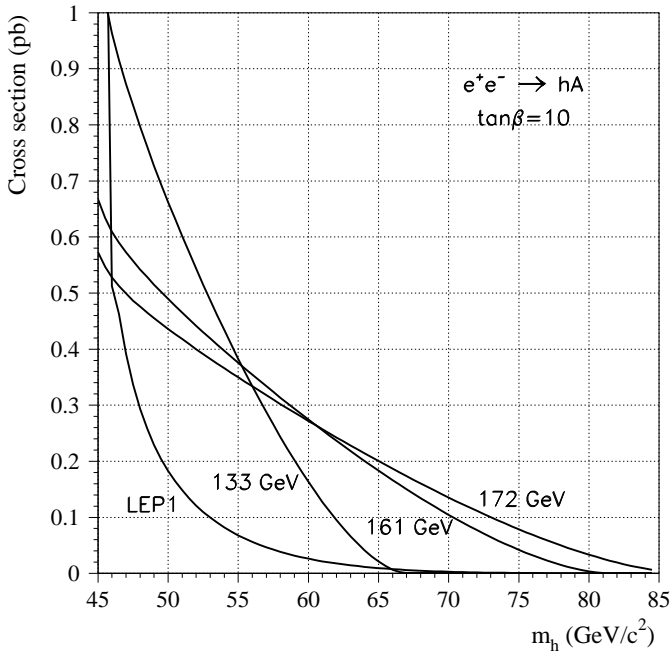}
  \end{center}
\end{minipage}
\hfill
\begin{minipage}{0.43\textwidth}
  \caption{Cross sections for $hA$ Higgs boson pair-production
           at various center-of-mass energies for $\tan\beta = 10$ 
           ($m_h\approx m_A$).}
  \label{fig:haxsec}
\end{minipage}
\vspace*{-0.5cm}
\end{figure}

In the Minimal Supersymmetric extension of the Standard Model 
(MSSM), Higgs boson masses and production cross
section are related. At the tree-level, only two free parameters 
describe the Higgs boson sector.
Typically, the parameter sets are $(m_{h,}m_{A})$, $(m_{h,}\tan \beta )$,
or $(m_{A,}\tan \beta )$. 
As an example, Fig.~\ref{fig:haxsec} (from~\cite{alephmssm}) 
gives the cross section for the reaction 
$e^{+}e^{-}\rightarrow hA$ for $\tan \beta = 10$ as a function of $m_{h}$.
Unlike for the MSM Higgs boson search, 161 and 172~GeV data are
almost equally important.

\begin{figure}[htbp]
\vspace*{-0.5cm}
\begin{minipage}{0.48\textwidth}
  \begin{center}
    \includegraphics[width=\textwidth]{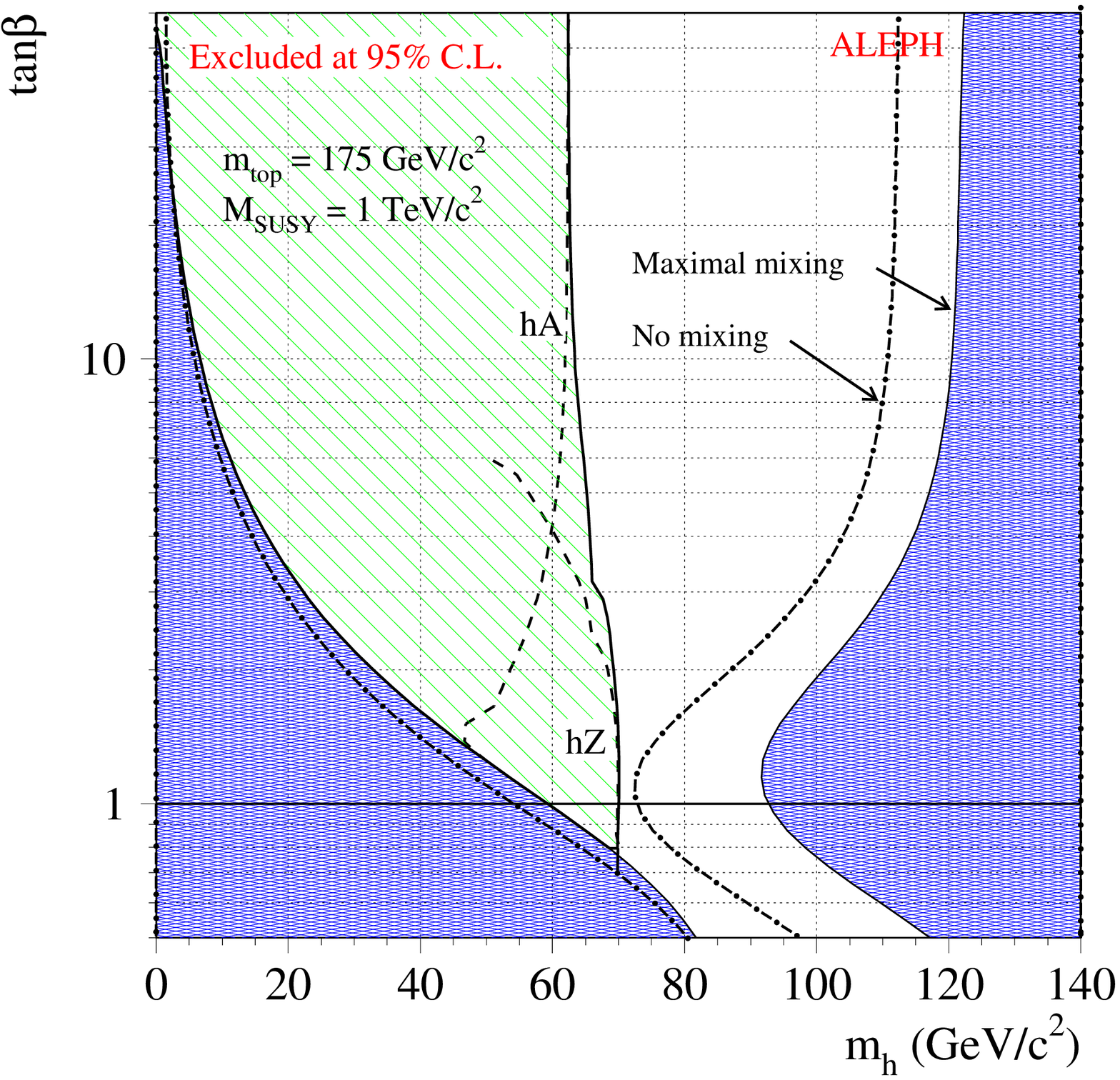}
  \end{center}
\vspace*{-0.5cm}
  \caption{\label{fig:mssmaleph}
           ALEPH MSSM exclusion for 161 to 172~GeV data. 
           The dark region is theoretically not allowed,
           and the hatched region is excluded. The dashed lines indicate the 
           ranges of exclusion from $hZ$ and $hA$ searches.}
\end{minipage}
\hfill
\begin{minipage}{0.48\textwidth}
  \begin{center}
   \includegraphics[width=\textwidth]{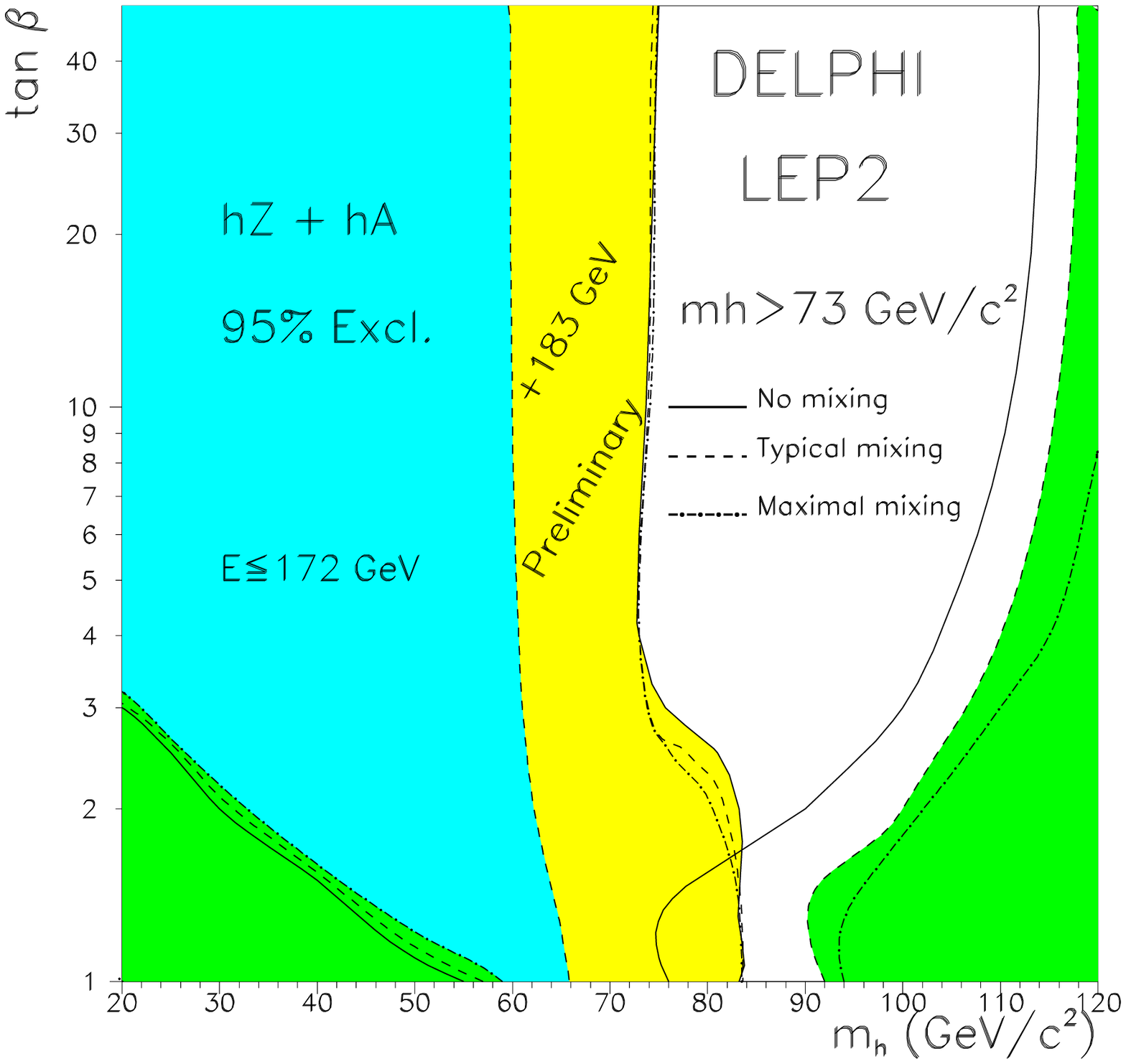}
  \end{center}
\vspace*{-0.6cm}
  \caption{\label{fig:mssmdel}
           DELPHI MSSM exclusion for 161 to 172~GeV data. 
           The dark region is theoretically not
           allowed, and the grey region is excluded.
           Preliminary results from 183~GeV data are included.}
\end{minipage}
\begin{minipage}{0.48\textwidth}
\vspace*{0.7cm}
  \begin{center}
    \includegraphics[width=\textwidth]{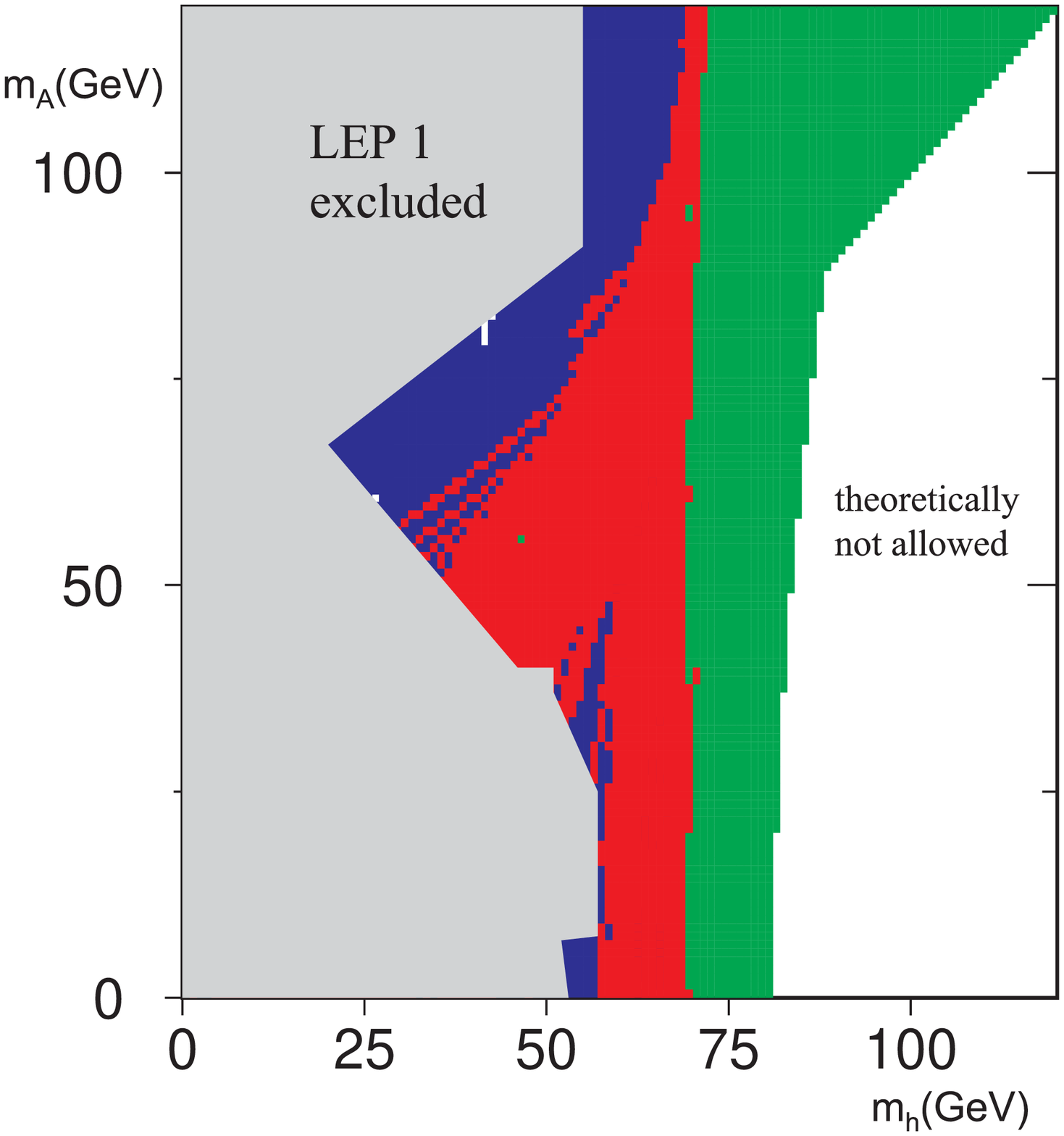}
  \end{center}
\vspace*{-1.3cm}
  \caption{\label{fig:172}
           Interpretation of 161 to 172~GeV DELPHI data.
           The region excluded by LEP1 (very light grey), 
           the newly 95\% CL excluded region at LEP2 (dark),
           the region where the exclusion depends on the SUSY parameter
           set (grey), the region with no sensitivity (light grey),
           and the theoretically not allowed region (white) are shown.}
\end{minipage}
\hfill
\begin{minipage}{0.48\textwidth}
\vspace*{0.7cm}
  \begin{center}
    \includegraphics[width=\textwidth]{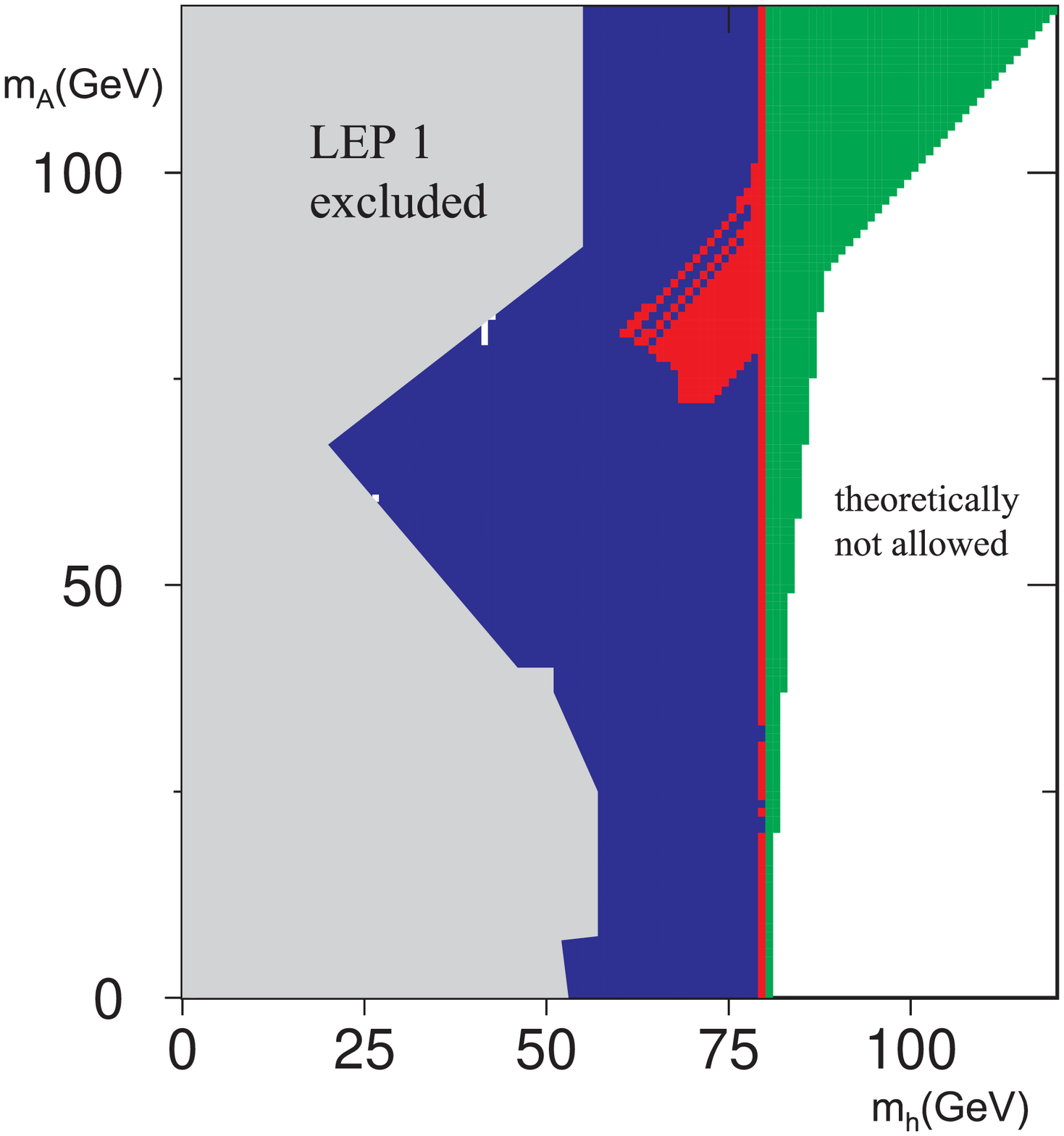}
  \end{center}
\vspace*{-1.3cm}
  \caption{\label{fig:172x4}
           Interpretation of 161 to 172~GeV data from
           four experiments.
           The region excluded by LEP1 (very light grey), 
           the newly 95\% CL excluded region at LEP2 (dark),
           the region where the exclusion depends on the SUSY parameter
           set (grey),
           the region with no sensitivity (light grey),
           and the theoretically not allowed region (white) are shown.}
\end{minipage}
\end{figure}

Interpretations of the ALEPH and DELPHI results are given in 
Figs.~\ref{fig:mssmaleph} (from~\cite{alephmssm}) 
and~\ref{fig:mssmdel} 
(from~\cite{delphi,lepc97}).
Three regions are shown: the excluded region at 95\% CL from the $%
e^{+}e^{-}\rightarrow hZ$ and $hA$ searches combined, the theoretically
disallowed region and the unexcluded region where a discovery is possible. 
The boundaries are given for three choices of 
mixing in the scalar top sector. This distinction
is made because of the important radiative correction effects which modify the
Higgs boson production rates. These radiative corrections are determined by
various other parameters of the MSSM, most importantly by the top and scalar 
top masses. The following (SUSY) parameter sets are used as 
proposed in~\cite{lep2}:
\vspace*{-2mm}
\begin{itemize}
\item  $m_{t}=175$~GeV, the top mass.
\item  $m_{sq}=1000$~GeV (also called $M_{\mathrm{SUSY}}$), the common
       mass parameter for all scalar quarks.
\item  $m_{g}=1000$~GeV, the gaugino mass.
\item  $\mu =-100$~GeV (no and typical mixing), and 
       $\mu =1000$~GeV (maximal mixing), the mixing parameter of the Higgs
       doublets in the MSSM superpotential.
\item  $A=0$ (no mixing), $A=1$ (typical mixing), and $A=\sqrt{6}$ (maximal
mixing), the mixing parameter in the scalar fermion sector, 
defined such\,that\,the\,mixing\,is\,proportional\,to\,$Am_{sq}$.
\end{itemize}
Note that for the low $\tan \beta $ region the $hZ$ searches, and for the
large $\tan \beta $ region the $hA$ searches determine the exclusion boundary.
Figure~\ref{fig:mssmdel} gives a preliminary bound $\tan\beta > 1.7$ 
independent of $m_h$ for the no-mixing case.
In the framework of the above parameter sets, 
ALEPH and DELPHI report preliminary results of
$m_h > 73$~GeV at 95\%~\cite{lepc97}.

In addition to the fixed sets of parameters, this article presents
an independent variation of the SUSY parameters.
Cancellation effects of production cross sections can occur
and thus some parameter regions are not excluded in this more 
general framework, as pointed out for LEP1 data in~\cite{jras}.
The SUSY parameters described above are varied 
in the following ranges:
\begin{itemize}
\item  $0.5 < \tan\beta < 50$.
\item  $200<m_{sq}<1000$~GeV.
\item  $200<m_{g}<1000$~GeV.
\item  $-500<\mu <500$~GeV.
\item  $-1<A<1.$
\end{itemize}
Each SUSY parameter combination defines the masses of neutralino, chargino and 
stop. Conservative experimental limits on these masses,
which exclude some parameter combinations, are taken into account.
No theoretical constraints on the MSSM are assumed.

The excluded regions are shown in Fig.~\ref{fig:172} in the 
$(m_{h,}m_{A})$ plane based on the DELPHI 161 to 172~GeV results.
The exclusion of an important region depends on the choice of the SUSY
parameters (central grey region).
In particular, no lower mass limit on either Higgs boson mass exists.
Examples of unexcluded SUSY parameter combinations leading to 
suppressed bremsstrahlung cross sections and large $m_h$ and $m_A$ mass 
differences are given in 
Table~\ref{tab:unexcl}\footnote{Other definitions in the literature are
$m_g = M_2 \leftrightarrow 0.5 M_2$, and $\mu \leftrightarrow -\mu$.}.

\begin{table}[hp]
\begin{center}
\begin{tabular} {|c|c|c|c|c|c|c|c|c|c|c|c|c|c|}\hline
$m_h$ &$m_A$&$m_t$ &$m_{sq}$&$m_g$&$\mu$ & $A$ & $\tan\beta$ & 
$m_{\tilde{t}1}$ & $m_{\tilde{t}2}$ & $\sigma_{hZ}^{161}$ &$\sigma_{hA}^{161}$ &
  $\sigma_{hZ}^{172}$ &$\sigma_{hA}^{172}$ \\ \hline
52.7  & 63  & 175  & 200    & 100 & $-500$  & 1   & 6           &
220              & 299              & 0.46 & 0.24 & 0.41 & 0.23   \\
74.5  & 12  & 175  & 1000   & 100 & $-500$  & 0   & 0.66        &
1001             & 1079             & 0.0 & 0.11 & 0.57 & 0.10   \\ \hline
\end{tabular}
\end{center}
\vspace*{-0.4cm}
\caption{\label{tab:unexcl}
Examples of unexcluded parameter combinations in the MSSM. 
Cross sections for Higgs boson bremsstrahlung and pair-production 
are given for $\surd s = 161$ and 172~GeV.
All masses are given in GeV and cross sections in pb.}
\end{table}

The same variation is repeated assuming four times the luminosity, which
corresponds approximately to the combined result of the four LEP experiments. 
Figure~\ref{fig:172x4} shows that the mass region where the exclusion
depends on the set of SUSY parameters is largely excluded at the 95\% CL
when the data are combined.
A lower mass limit of about 60~GeV on the CP-even Higgs boson is set,
while no mass limit on the CP-odd Higgs boson exists
(for small $m_A$ values, unexcluded parameter combinations 
exist for $0.5 < \tan\beta < 1$).

\section{\label{sec:outlook} Prospects}

\begin{figure}[htbp]
\begin{minipage}{0.48\textwidth}
\vspace*{-0.2cm}
  \begin{center}
    \includegraphics[width=\textwidth]{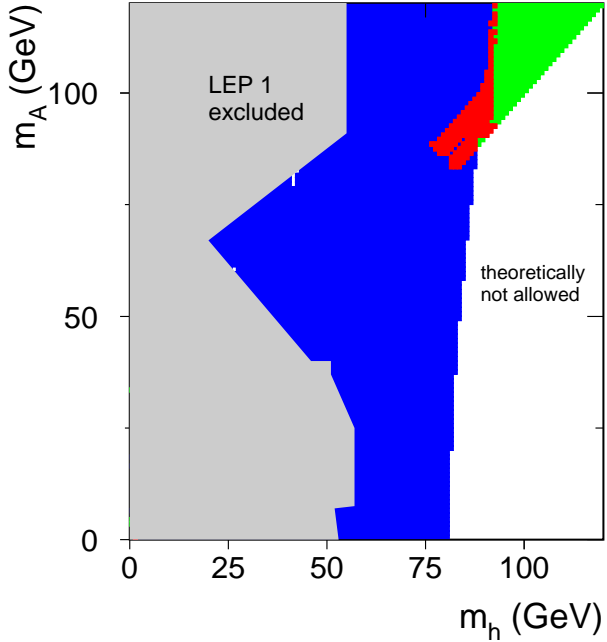}
  \end{center}
\end{minipage}
\hfill
\begin{minipage}{0.48\textwidth}
\vspace*{-0.5cm}
  \caption{\label{fig:182} MSSM prospects 
           for $\surd s = 183$~GeV and ${\cal L} = 200$~pb$^{-1}$.
           The region excluded by LEP1 (very light grey), 
           the 95\% CL sensitivity region at LEP2 (dark),
           the region where the sensitivity depends on the SUSY parameter
           set (grey), 
           the region with no sensitivity (light grey),
           and the theoretically not allowed region (white) are shown.}
\end{minipage}
\vspace*{-0.4cm}
\end{figure}

Previously, detailed studies have been presented for the preparation
of the LEP2 run and sensitivity ranges have been given~\cite{lep2}.
In order to estimate the sensitivity reach of the current LEP2 run at
183~GeV, the SUSY parameter scan is repeated,
assuming the same experimental performance and a total luminosity
of 200~pb$^{-1}$, corresponding to 50~pb$^{-1}$ for each LEP experiment.
The MSSM prospects are given in Fig.~\ref{fig:182}.
For large $m_A$ the limit on $m_h$ is equal to the limit on the MSM Higgs 
boson.
Independent of the SUSY parameter choice,
lower mass limits on CP-even and CP-odd Higgs bosons could be set.
These limits are about 60 and 73~GeV, respectively. 
For the combined data of the four LEP experiments these limits are 
about 76 and 83~GeV. 
The discovery reach is slightly lower, assuming
a $5\sigma$ discovery effect, while the 95\% CL exclusion
corresponds to a $2\sigma $ effect. For the combined sensitivity of the four
LEP experiments, a discovery of an 85~GeV MSM Higgs boson or the
exclusion of a 90~GeV Higgs boson is anticipated with the 1997
data.

It is exciting that a further energy increase to about 190~GeV is
planned for 1998, with the potential to find a Higgs boson with a mass 
of about 95~GeV. Particular larger data statistics are needed to 
find the  Higgs boson near the Z boson mass.
A possible further energy increase to about 200~GeV at a
later stage is discussed, which is strongly motivated in the MSSM by the 
fact that the lightest Higgs boson can be found for $\tan \beta < 2$, 
even for unfavorable values of the unknown SUSY parameters.

The sensitivity mass range for a charged Higgs boson 
depends largely on the total integrated luminosity, and 
will extend to about 70~GeV for 
${\cal L} = 500$~pb$^{-1}$~\cite{as94,lep2}.

\section{Conclusions}
No Higgs boson signal has been observed.
The Minimal Standard Model Higgs boson mass limit at the 95\%~CL
is 77.5~GeV for combined 1996 data, and individual mass limits of up to
88.6~GeV are reported for 1997 data.
Much progress has also been made in the searches for neutral and
charged Higgs boson pair-production. Constraints on the parameters
of the Minimal Supersymmetric extension of the Standard Model
have significantly improved. The combination of the results from 
the four LEP experiments is and will be of great importance to 
significantly increase the discovery sensitivity and to determine the
excluded mass regions.
The LEP experiments have an excellent potential for a discovery 
during the next three years.

\vspace*{0.6cm}
\noindent{\Large \bf Acknowledgements}
\vspace*{0.3cm}

I would like to thank my fellow Higgs hunters for many fruitful 
discussions, the organizers of the Moscow workshop for their 
warm hospitality, the Aspen Center for Physics for their valuable
support, and Wim de Boer for advice on the manuscript.

\newpage

\def\ee{${e^+ e^-}$ }
\newcommand{\sqrts}{\mbox{$\sqrt {s}$}}


\begin{thebibliography}{99}

\bibitem{eilam}
E.~Gross, {\it Title Search for New Physics at LEP 2 (ECM $=$ 130--172 GeV)}, 
talk at the Moriond Conference 1997, 
to be published in the proceedings.

\bibitem{bill} 
W.~Murray, {\it Title Search for the Standard Model Higgs boson at LEP}, 
talk at the HEP'97 Conference, Jerusalem,
Aug. 19--26, 1997, to be published in the proceedings.

\bibitem{patrick} 
P. Janot, {\it Searches for New Particles},
talk at the HEP'97 Conference, Jerusalem,
Aug. 19--26, 1997, to be published in the proceedings.

\bibitem{as97}
A.~Sopczak, 
{\it Higgs Bosons Searches at LEP},
DESY97-129, ISSN 0418-9833, 
Proc. XII Int. Symposium on High Energy Physics, Gauhati, India,
Dec. 1996 -- Jan. 1997.

\bibitem{karlsruhe}
W.~de~Boer, 
{\it In Search of SUSY},
{\it Acta Phys. Pol.} {\bf B 28} (1997) 1395.

\bibitem{lepew}
D.~Ward, 
{\it Test of the Standard Model $W$ mass and $WWZ $coupling}, 
talk at the HEP'97 Conference, Jerusalem, Aug. 19--26, 1997, 
to be published in the proceedings; \\
J. Erler and P. Langacker,
{\it The 1997 off-year partial update for the 1998 edition of Particles Data},
available on the PDG Web-pages (URL: http://pdg.lbl.gov/).

\bibitem{jens}
J.~Erler, Private communications.

\bibitem{aleph} ALEPH Coll., R. Barate et al., 
{\it Search for the Standard Model Higgs Boson in \ee Collisions at 
\sqrts\ = 161, 170 and 172~GeV}, CERN-PPE/97-070,
to be published in Phys.\,Lett.\,{\bf B}.

\bibitem{delphi} DELPHI Coll., P. Abreu et al., {\it Search for
neutral and charged Higgs bosons in \ee 
collisions at \sqrts\ = 161 and 172~GeV},
CERN-PPE/97-85, to be published in Z.\,Phys.\,{\bf C}.

\bibitem{lepwg} 
ALEPH, DELPHI, L3, and OPAL Coll., and the LEP Working Group for Higgs
Boson Searches,
{\it Lower bounds for the SM Higgs Boson mass: combined result from the 
     four LEP experiments},
CERN/LEPC 97-11, Nov 3, 1997.

\bibitem{l3} L3 Coll., M. Acciarri et al., 
{\it Search for the Standard Model Higgs Boson in \ee\ Interactions
at 161 $\leq$ \sqrts\ $\leq$ 172~GeV}, CERN-PPE/97-81,
to be published in Phys.\,Lett.\,{\bf B}.

\bibitem{opal} 
OPAL Coll., K. Ackerstaff et al.,
{\it Search for the Standard Model Higgs Boson in 
\ee Collisions at \sqrts\ = 161, 170 and 172~GeV}, CERN-PPE/97-115, 
to\,be\,published\,in\,Z.Phys.{\bf C}.

\bibitem{lepc97}
LEP Experiment Committee OPEN SESSION
{\it Reports on the LEP Experiments: ALEPH, DELPHI, L3, OPAL},
CERN, Nov. 11, 1997.

\bibitem{jadib} 
ALEPH method: P. Janot and F. Le Diberder, 
{\it Combining `Limits'},
CERN-PPE/97-053 and LPNHE 97-01, to be published in Nucl. Instrum. Methods.

\bibitem{alex}
DELPHI method: A.L.~Read, {\it Optimal statistical analysis of search 
results based on the likelihood ratio and its application to the search 
for the MSM Higgs boson at 161 and 172 GeV}, DELPHI note 97-158 PHYS 737,
Oct. 1997.

\bibitem{l3method} 
L3 method: A. Favara and M. Pieri, {\it Confidence level 
estimation and analysis optimisation}, Firenze University preprint 
DFF-278/4/1997, E-preprint hep-ex 9706016.

\bibitem{bock} 
OPAL method: P. Bock, {\it Determination of exclusion limits for
particle production using different decay channels with different efficiencies,
mass resolutions and backgrounds}, Heidelberg University preprint HD-PY-96/05
(1996), to\,be\,published\,in Nucl.\,Instrum.\,Methods.

\bibitem{PDG}  
Particle Data Group: R. M. Barnett et al., {\it Review of
Particle Physics}, Phys. Rev. {\bf D 54} (1996) 1.

\bibitem{blondel}
A.~Blondel, 
{\it What can precision electroweak data tell us about electroweak
symmetry breaking?}, to appear in {\it Higgs Physics}, editor G.L.~Kane,
published by World Scientific.

\bibitem{alephmssm}
ALEPH Coll., R. Barate et al.,
{\it Search for the neutral Higgs bosons of the MSSM in \ee collisions at 
$\sqrt{s}$ from 130 to 172 GeV},
CERN-PPE/97-071, to\,be\,published\,in\,Phys.\,Lett.\,{\bf B}.

\bibitem{delphihphm}
DELPHI Coll., P. Abreu et al.,
{\it Search for charged Higgs bosons in \ee 
collisions at $\sqrt{s} = 172 $~GeV},
CERN-PPE/97-145, to be published in Phys. Lett. {\bf B}.

\bibitem{alephhphm}
ALEPH Coll., R. Barate et al.,
{\it Search for charged Higgs-Bosons in \ee collisions at centre-of-mass energies from 130 to 172 GeV}, 
CERN-PPE/97-129, to be published in Phys.\,Lett.\,{\bf B}.

\bibitem{l3hphm}
L3 Coll., O.~Adriani et al., 
{\it Search for Non-Minimal Higgs Bosons in Z0 Decays},
Z. Phys. {\bf C 57} (1993) 355.
 
\bibitem{opalhphm}
OPAL Coll., K. Ackerstaff et al.,
{\it Search for Charged Higgs Bosons Pair-Produced 
in \ee Collisions at $\sqrt{s}=130-172$ GeV}, 
contributed paper HEP'97 Conference, Jerusalem, Aug. 19--26, 1997.

\bibitem{noncp}
DELPHI Coll., C.~Martinez-Rivero,
{\it Search for Neutral Higgs Bosons in Two-Doublet Models at LEP1 and
LEP2 Energies}, 
DELPHI note 97-82 CONF 68, July 20, 1997, contributed paper HEP'97 Conference,
Jerusalem, Aug. 19--26, 1997.

\bibitem{jack}
J.F.~Gunion, B.~Grzadkowski, H.E.~Haber and J.~Kalinowski, 
{\it LEP Limits on CP-Violating Non-Minimal Higgs Sectors},
UCD-97-11, April 1997.

\bibitem{l3inv}
L3 Coll., M. Acciarri et al.,
{\it Missing mass spectra in hadronic events from \ee collisions at 
$\sqrt{s} = 161-172$~GeV and limits on invisible Higgs decays},
CERN-PPE/97-97, 24 July 1997, to be published in Phys. Lett. {\bf B}.

\bibitem{opalhgamma}
OPAL Coll., K. Ackerstaff et al.,
{\it Search for a Massive Di-photon Resonance at $\sqrt{s} = 91$ -- 172 GeV},
CERN-PPE/97-121, to be published in Z. Phys. {\bf C}.

\bibitem{lep2}
E.~Accomando et al., ``Higgs Physics'' in Physics at LEP2, 
CERN 96-01 (1996) p.~351.

\bibitem{jras}
J.~Rosiek and A.~Sopczak, 
{\it Present and future searches with \ee colliders 
for the neutral Higgs bosons of the Minimal Supersymmetric 
Standard Model -- the complete 1-loop analysis},
{Phys. Lett.} {\bf B 341} (1995) 419.

\bibitem{as94}
A.~Sopczak, 
{\it Higgs boson discovery potential beyond the minimal standard model},
{Int. J. Mod. Phys.} {\bf A 9} (1994) 1747.

\end{thebibliography}
\end{document}